\documentclass{SciPost}
\usepackage{orcidlink}

\hypersetup{
    colorlinks,
    linkcolor={red!50!black},
    citecolor={blue!50!black},
    urlcolor={blue!80!black}
}

\usepackage[bitstream-charter]{mathdesign}
\usepackage{soul}
\usepackage[normalem]{ulem}
\usepackage[bottom]{footmisc}

\DeclareSymbolFont{usualmathcal}{OMS}{cmsy}{m}{n}
\DeclareSymbolFontAlphabet{\mathcal}{usualmathcal}

\fancypagestyle{SPstyle}{
\fancyhf{}
\lhead{\colorbox{scipostblue}{\bf \color{white} ~SciPost Physics }}
\rhead{{\bf \color{scipostdeepblue} ~Submission }}

\fancyfoot[C]{\textbf{\thepage}}
}

\newcommand{\edit}[1]{\textcolor{black}{#1}}

\newcommand{\ms}[0]{_{\text{m}}}
\newcommand{\tp}[0]{_{\text{tp}}}
\newcommand{\cb}[0]{_{\text{c}}}

\begin{document}

\pagestyle{SPstyle}

\begin{center}{\Large \textbf{\color{scipostdeepblue}{
Instanton theory and fluctuation corrections to the thermal nucleation rate of a ferromagnetic superfluid
}}}\end{center}

\begin{center}\textbf{
Enrique Rozas Garcia\,\orcidlink{0000-0002-6483-0228}\textsuperscript{1$\star$} and
Johannes Hofmann\,\orcidlink{0000-0002-0667-2452}\textsuperscript{1,2$\dagger$}
}\end{center}

\begin{center}
{\bf 1} Department of Physics, Gothenburg University, Sweden
\\
{\bf 2} Nordita, Stockholm University and KTH Royal Institute of Technology, Stockholm, Sweden
\\[\baselineskip]
$\star$ \href{mailto:enrique.rozas.garcia@physics.gu.se}{\small enrique.rozas.garcia@physics.gu.se}\,,\quad
$\dagger$ \href{mailto:johannes.hofmann@physics.gu.se}{\small johannes.hofmann@physics.gu.se}
\end{center}

\section*{\color{scipostdeepblue}{Abstract}}
\textbf{\boldmath{We provide a field-theoretical description of thermal nucleation in a one-dimensional ferromagnetic superfluid, a quantum-gas analogue of false-vacuum decay. The rate at which stable-phase domains nucleate follows an Arrhenius law, with an exponential factor determined by a saddle point of the energy functional---the critical droplet or instanton---and a prefactor fixed by fluctuations about this configuration. We evaluate both contributions over the full parameter space, using a Gel'fand-Yaglom approach to reduce the calculation of the fluctuation contribution to an initial-value problem. In addition, we obtain a closed-form expression for critical droplets close to the phase coexistence line, and use it to formulate an effective theory of domain nucleation and growth as a Kramers escape problem for the domain-wall separation. Our results determine the parametric dependence of the nucleation rate and predict its signature in experimental images of a nucleating gas.
}}

\vspace{\baselineskip}



\vspace{10pt}
\noindent\rule{\textwidth}{1pt}
\tableofcontents
\noindent\rule{\textwidth}{1pt}
\vspace{10pt}


\section{Introduction}

A system that is quenched across a first-order phase transition into a metastable state relaxes to equilibrium via the nucleation and growth of stable-phase domains. Originally described in the context of liquid-gas transitions \cite{gibbs1878equilibrium}, nucleation is ubiquitous in nature and industry: Ice crystals, water droplets, and other aerosols are familiar examples of nucleation in the atmosphere \cite{laaksonen1995nucleation, zhang2012nucleation}, and the industrial production of chemicals and pharmaceuticals commonly involves the nucleation of crystals in solutions \cite{wu2022nucleation}. On a cosmological scale, it has been suggested that first-order phase transitions in the early universe may generate a background of gravitational waves that carry information from these early times \cite{kosowsky1992gravitational, hindmarsh2021phase}. Direct detection of cosmological signals remains challenging, however, and table-top experiments using cold-atom analogues have emerged as a platform to probe nucleation in a controlled environment~\cite{viermann2022quantum, zhu2024probing, zenesini2024false, cominotti2025observation, vodeb2025stirring, osterholz2025collective, luo2025quantum}. The aim of this paper is to present a systematic theoretical calculation of the nucleation rate in ferromagnetic superfluid quantum gases---as probed experimentally in~\cite{zenesini2024false,cominotti2025observation}---and to provide a model for nucleation that can guide further measurements.

The theory of nucleation traces back to the work of Gibbs on the thermodynamic stability of heterogeneous mixtures \cite{gibbs1878equilibrium}. Volmer and Weber \cite{volmer1926keimbildung} and Becker and D\"oring \cite{becker1935kinetische} later derived Arrhenius-like expressions for the rate of nucleation, and Zeldovich \cite{zeldovich1943theory} established a connection with the work of Kramers \cite{kramers1940brownian} on stochastic barrier-crossing. Their combined works form the core of Classical Nucleation Theory~\cite{kelton2010nucleation}, a phenomenological theory for extended systems trapped by an energy barrier in a metastable state (or \textit{false vacuum} in cosmological terminology). In Classical Nucleation Theory, thermal fluctuations stochastically create droplets of the stable phase (or \textit{true vacuum}); small droplets shrink and disappear due to their high surface tension, while large droplets grow as their energetically favorable bulk overcomes the interfacial energy cost. This competition between interface and bulk defines a critical size that separates shrinking from growing droplets. The rate of nucleation is given by the probability of thermal fluctuations creating such critical droplets.

The classical assumption of clearly separated bulk and interfacial energy contributions breaks down when the size of a droplet is comparable to the thickness of its interface. This led to the development of nucleation field theories by Cahn and Hilliard \cite{cahn1959free} and Langer \cite{langer1967theory, langer1969statistical, langer1974metastable, langer1980kinetics}, where the critical droplet emerges as a saddle point of the free energy functional. Here, the nucleation rate $\Gamma$ is defined as the steady-state probability flux through said saddle point and splits into two contributions:

\begin{align}
    \Gamma = \frac{\sigma }{\pi} \, \left|\text{Im} (\ln \mathcal Z)\right| = \frac{\sigma}{\pi} \times \frac{1}{2} D^{-\frac{1}{2}} \; L\left(\frac{\beta \Delta E}{2\pi}\right)^{\frac{1}{2}} \; \, e^{-\beta \Delta E}.
    \label{eq:l_rate}
\end{align}
\edit{The first contribution contains a dynamical prefactor $\sigma$, which describes the exponential growth rate \mbox{$\sim e^{\sigma t}$} of small perturbations to the critical droplet.} The second contribution, set by the partition function~${\cal Z}$, is the probability of observing a critical droplet as the result of an equilibrium fluctuation in the metastable phase. This equilibrium contribution has, in turn, three factors written out on the right-hand side of Eq.~\eqref{eq:l_rate}: First, an exponential Arrhenius-like term that depends on the inverse temperature $\beta$ and the energy barrier preventing nucleation $\Delta E$ (i.e., the energy difference between the critical droplet and the metastable state). Second, a temperature-dependent prefactor proportional to the system size $L$, which arises from translational invariance. Finally, Eq.~\eqref{eq:l_rate} contains a fluctuation determinant $D$, which characterizes the contribution of small deformations of the critical droplet to the nucleation rate.

Equation~\eqref{eq:l_rate} is valid when the typical energy of thermal fluctuations is smaller than the energy barrier. However, close to absolute zero temperature, thermal fluctuations become less relevant and quantum tunneling becomes the leading contribution to the nucleation rate. Here, Coleman and Callan~\cite{coleman1977fate, callan1977fate} developed a zero-temperature description of tunneling quantum fields to describe cosmological false-vacuum decay, with a tunneling rate
\begin{equation}
    \Gamma_Q = \frac{2}{\hbar} \left|\text{Im}(E_0)\right|, \label{coleman_rate}
\end{equation}
where $E_0$ is the ground-state energy of the metastable field. While the transition between quantum and thermal behavior \cite{affleck1981quantum, linde1983decay, hanggi1990reaction} has been observed in systems with a single degree of freedom \cite{devoret1985measurements}---receiving the 2025 Nobel Prize in Physics---it is still an open challenge for spatially extended systems.

\edit{Indeed, while numerical simulations \cite{pirvu2024thermal, pirvu2026thermal, hirvonen2026langer} are consistent with the theory of nucleation, a quantitative experimental confirmation remains a topic of debate \cite{langer1980kinetics, oxtoby1992homogeneous, ryu2010validity, nellas2010exploring, hindmarsh2024ab}. This is typically attributed to difficulties in isolating experiments from impurities and external influences, or in determining the correct theoretical model for a given setup. A striking example is the A-B transition in superfluid $^3$He, where the nucleation time is predicted to be longer than the lifetime of the universe, yet is experimentally measured in hours \cite{leggett1984nucleation, schiffer1992strong, hindmarsh2024ab}.} In recent years, quantum gases have thus emerged as an experimental platform to study nucleation \cite{zhu2024probing, zenesini2024false, cominotti2025observation, vodeb2025stirring, osterholz2025collective, luo2025quantum} as they offer clean systems, precise experimental control, and well-established microscopic theoretical models \cite{fialko15, lagnese2021false, maki2023monte, lagnese2023detecting, maertens2025real, johansen2025many}. 

In particular, the experiments by Zenesini {\it et al.}~\cite{zenesini2024false} and Cominotti {\it et al.}~\cite{cominotti2025observation} measure the rate of nucleation for a one-dimensional (1D) atomic superfluid with two internal states---up~($\uparrow$) and down~($\downarrow$)---that are coherently coupled by microwave radiation. Such systems are described by the dimensionless mean-field energy functional~\cite{cominotti2023ferromagnetism, zenesini2024false}\footnote{We express distances in units of the healing length \mbox{$\xi = (\hbar/2m|\kappa|n)^{1/2}$} and energy in units of~$\hbar n^2\xi |\kappa|/2$ to obtain the energy functional~\eqref{eq:energy} without additional prefactors. Here, $m$ is the atomic mass, \mbox{$n = n_{\uparrow} + n_\downarrow$} the constant condensate density, and \mbox{$\kappa=(g_{\downarrow\downarrow}+g_{\uparrow\uparrow})/(2\hbar)-g_{\downarrow\uparrow}/\hbar<0$} contains the intracomponent ($g_{\downarrow\downarrow}, g_{\uparrow\uparrow}$) and intercomponent ($g_{\downarrow\uparrow}$) interaction constants. \edit{Note that $\beta$ has units of inverse energy, i.e., of~$2/\hbar n^2\xi |\kappa|$.} \label{footnote:1}}
\begin{equation}
    E[\theta(x)] = \int_{-\infty}^{\infty} dx \, \left\{ \frac{1}{2}\left(\frac{d\theta}{dx}\right)^2 + V[\theta]\right\} \label{eq:energy}
\end{equation}
with the potential
\begin{equation}
    V[\theta] = a\sin\theta -\frac{\sin^2\theta}{2} - c\cos\theta.
    \label{eq:V}
\end{equation}
The polarization field $\theta(x)$ characterizes the local population imbalance, \mbox{$\sin\theta = (n_{\uparrow} - n_\downarrow)/n$}, where $n_\uparrow$ and $n_\downarrow$ are the densities of the two internal states and \mbox{$n=n_\uparrow+n_\downarrow$} is the total density. The dimensionless parameters \mbox{$a = -\delta_f/|\kappa|n$} and \mbox{$c=\Omega_R/|\kappa|n$} depend on the experimentally tunable Rabi frequency $\Omega_R$ and detuning $\delta_f$. Equation~\eqref{eq:V} describes a double-well potential with tilt controlled by the parameter~$a$. This potential is analogous to the free energy of a homogeneous ferromagnetic system, where energy minima represent different magnetization states and the tilt is induced by an externally applied magnetic field \cite{cominotti2023ferromagnetism}. In the experiments~\cite{zenesini2024false, cominotti2025observation}, after equilibration to an initial stable state, the detuning $a$ is adiabatically increased to transfer the system into a homogeneous metastable state. Spatially resolved imaging of the density distribution for multiple realizations and times after the quench allows the direct observation of domains as they nucleate. Fitting the average nucleated fraction to a single-time-scale empirical function yields a nucleation rate, and an additional fit to an Arrhenius-like law yields a temperature and prefactor. The temperatures obtained indicate that nucleation is driven by thermal, rather than quantum, fluctuations \cite{cominotti2025observation}.

A quantitative experimental confirmation of the thermal field theory of nucleation requires a full evaluation of the rate~\eqref{eq:l_rate}, whereas current experimental results are analyzed without the fluctuation prefactor and using approximations of the exponential term. In this paper, we perform the full calculation for a ferromagnetic superfluid described by the mean-field energy~\eqref{eq:energy}. In Sec.~\ref{sec:theory}, we show how the different contributions to the nucleation rate arise from a saddle-point evaluation of the field-theoretical partition function. We determine the exponential contribution from the critical droplet and evaluate the prefactor over the full parameter space. Our results clarify the dependence of the different contributions to the nucleation rate on the parameters of the system, which reduces the sources of systematic error for future experimental fits. Furthermore, we provide analytical results for the shape of the critical droplet for quenches close to the first-order phase boundary. In Sec.~\ref{sec:Kramers}, we use these analytical results to formulate an effective theory for the nucleation process as a Kramers escape problem. This effective theory leads to a time evolution of the average population imbalance that includes both the nucleation and growth rates of droplets as two distinct time scales, whereas previous analyses used a single relevant time scale. We conclude in Sec.~\ref{sec:conclusions} with a summary of our results and an outlook on possible extensions, e.g., to other energy functionals and higher dimensions. Technical details are provided in three Appendices.

\section{Field theory of metastable decay}
\label{sec:theory}

This section evaluates the nucleation rate~\eqref{eq:l_rate} for a ferromagnetic superfluid described by the energy functional~\eqref{eq:energy}. We determine the critical droplet and its energy in Sec.~\ref{subsec:inst} and compute the rate prefactor in Sec.~\ref{subsec:prefactor}. The full nucleation rate is discussed in Sec.~\ref{subsec:result}.

\begin{figure}[t]
    \centering
    \includegraphics[width=\linewidth]{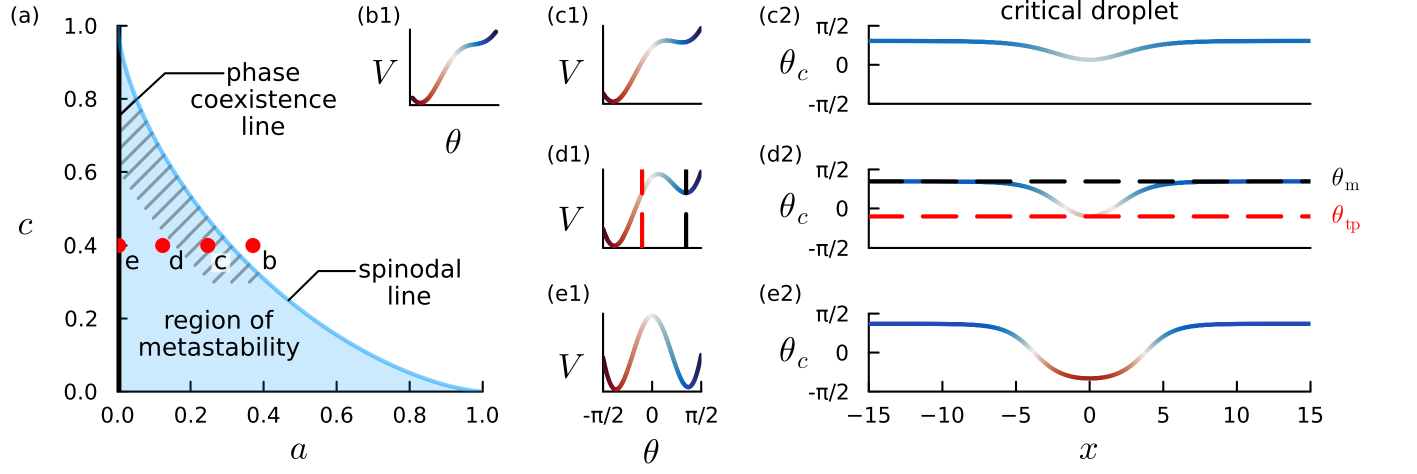}
    \caption{(a) Phase diagram for the mean-field potential $V[\theta]$~\eqref{eq:V} as a function of dimensionless detuning $a$ and Rabi frequency $c$. The metastability region is shaded in blue; the hatched portion indicates the parameter range explored in the experiment~\cite{zenesini2024false}. (b1-e1) Potential~$V[\theta]$ evaluated at the red points in panel~(a), illustrating the effect of increasing the tilt from almost-degenerate minima~(e1), to decreasing potential barriers (d1,c1), until the metastable state vanishes~(b1). (c2-e2) Critical-droplet configurations associated with the points (c,d,e); there is no critical droplet associated with (b1) as it lies outside the metastability region. The magnitude of the polarization field $\theta$ is represented as a gradient from red ($-\pi/2$) to blue ($\pi/2$).}
    \label{fig:1}
\end{figure}

Figure~\ref{fig:1} summarizes the basic elements of the theory. Panel~(a) shows the phase diagram of the mean-field potential $V[\theta]$ as a function of the dimensionless detuning~$a$ and the dimensionless Rabi frequency~$c$. The blue area indicates the parameter region with a positive-$\theta$ homogeneous metastable phase, i.e., the region where $V[\theta]$ has both a global and a local minimum [see Figs.~\ref{fig:1}(c1)-(e1)]. \edit{To the left, the metastability region is bounded by the phase coexistence line, \mbox{$a=0$} [bold black line in Fig.~\ref{fig:1}(a)], where the potential is symmetric and has two degenerate minima separated by a large energy barrier [Fig.~\ref{fig:1}(e1) shows \mbox{$a\approx 0$}]. Increasing the value of $a$ tilts the potential, which raises the energy of the metastable phase and thus lowers the height of the barrier [Figs.~\ref{fig:1}(d1,c1)]. To the right, the metastability region is bounded by the spinodal line [light blue line in Fig.~\ref{fig:1}(a)]
\begin{equation}
    a_c = (1-c^{2/3})^{3/2}. \label{eq:ac}
\end{equation}
As \mbox{$a\to a_c$}, the barrier height approaches zero. For tilts beyond the spinodal line, \mbox{$a>a_c$}, the potential has a single minimum [Fig.~\ref{fig:1}(b1)]. The phase coexistence and spinodal lines meet at the second-order critical point \mbox{$(a,c)=(0,1)$}.}

The equilibrium contribution to the nucleation rate \eqref{eq:l_rate} is determined by the energy landscape around two distinguished field configurations: the homogeneous metastable state $\theta\ms$---the initial configuration of the field and a local minimum of the energy functional~\eqref{eq:energy}---and the critical droplet $\theta\cb(x)$, which represents the tipping point between shrinking and growing droplets, i.e., a saddle point of the energy functional. These two configurations dominate the partition function in the low-temperature limit,
\begin{equation}
    \label{eq:Z}
    \mathcal{Z} = \int \mathcal{D}[\theta(x)] \, e^{-\beta E[\theta(x)]} \approx \mathcal{Z}\ms + \edit{\frac{i}{2}}\left|\mathcal{Z}\cb\right|.
\end{equation}
Here, $\mathcal{D}[\theta]$ is the path integral measure over the field $\theta(x)$ and $\mathcal{Z}_{\rm sp}$ is the saddle-point contribution of the metastable state (sp$=$m) and critical droplet (sp$=$c). \edit{The factor $1/2$ arises from the analytic continuation of the critical-droplet unstable mode~\cite{langer1974metastable,callan1977fate}. Physically, it reflects that only one of the two directions along the unstable mode leads to a growing droplet, while the other returns to the metastable state.}

We parameterize fluctuations around the stationary-energy points by \mbox{$\theta(x) = \theta_{\rm sp}(x) + \delta \theta(x)$}. By the equipartition theorem, the amplitude of the fluctuations is expected to be small (of order $1/\sqrt{\beta}$), such that the energy~\eqref{eq:energy} may be expanded to second order in $\delta \theta$. This gives
\begin{equation}
    \mathcal{Z}_{\rm sp} = e^{-\beta E_{\rm sp}} \int \mathcal{D}[\delta\theta]\; \exp\left[-\frac{\beta}{2}\int_{-\infty}^{\infty} dx\; \delta\theta(x) \, \mathcal{K}_{\rm sp} \, \delta\theta(x)\right] = \frac{e^{-\beta E_{\rm sp}}}{\sqrt{\det\mathcal{K}_{\rm sp}}}, \label{eq:Z0_temp1}
\end{equation}
where \mbox{$E_{\rm sp}=E[\theta_{\rm sp}(x)]$}, and we have formally evaluated the Gaussian integral as the determinant of the fluctuation operator
\begin{equation}
     \mathcal{K}_{\rm sp} = -\frac{d^2}{dx^2} + \frac{\partial^2 V}{\partial \theta^2} \biggr|_{\theta_{\rm sp}(x)}, \label{eq:K}
\end{equation}
which describes the energy cost of deforming $\theta_{\rm sp}(x)$. As detailed in Sec.~\ref{subsec:prefactor}, all eigenvalues of $\mathcal{K}\ms$ are positive while the critical-droplet fluctuation operator $\mathcal{K}\cb$ possesses a single negative eigenvalue, $\lambda_{-1}$, which yields the imaginary contribution to Eq.~\eqref{eq:Z}. The operator $\mathcal{K}\cb$ also possesses a single zero mode associated with translation symmetry, which gives a contribution to $\mathcal{Z}\cb$ proportional to the system size $L$.

With this, we obtain the equilibrium contribution to the nucleation rate, Eq.~\eqref{eq:l_rate}, as
\begin{equation}
    \left|\text{Im}\left(\ln\mathcal{Z}\right)\right| 
    = \edit{\frac{1}{2}}\frac{\left|\mathcal{Z}\cb\right|}{\mathcal{Z}\ms} = \edit{\frac{1}{2}}\frac{1}{\sqrt{D(a, c)}} \, L \left(\frac{\beta \Delta E}{2\pi}\right)^{1/2} \; e^{-\beta \Delta E}, \label{eq:Zratio}
\end{equation}
where we expand in \mbox{$|\mathcal{Z}\cb| \ll \mathcal{Z}\ms$}. Here, $\Delta E$ is the energy difference between the critical droplet and the metastable state,
\begin{align}
    \Delta E = E\cb - E\ms = E[\theta\cb(x)] - E[\theta\ms],
    \label{eq:Ediff}
\end{align}
which we evaluate in the next section [Eq.~\eqref{eq:deltaE_full}]. The fluctuation determinant is given by
\begin{equation}
    D(a, c) = \frac{|\det'\mathcal{K}\cb|}{\det\mathcal{K}\ms}, \label{eq:D}
\end{equation}
where $\det'$ denotes the omission of the above-mentioned zero eigenvalue of $\mathcal{K}\cb$ from the determinant. Section~\ref{subsec:prefactor} contains a detailed derivation and evaluation of the terms in the prefactor.

\edit{
The dynamical prefactor $\sigma$ in Eq.~\eqref{eq:l_rate} is determined by the exponential growth of the unstable critical droplet. A linear-stability analysis\footnote{The nucleation rate in Eq.~\eqref{eq:l_rate}  implicitly assumes the stochastic evolution equation
\begin{equation}
    m\frac{\partial^2 \theta}{\partial t^2} + \gamma\frac{\partial \theta}{\partial t} = - \frac{\delta E[\theta]}{\delta \theta} + \eta(x,t), \label{eq:langer_dynamics}
\end{equation} 
where $\eta(x,t)$ is a spatiotemporal white noise with variance fixed by the fluctuation-dissipation theorem \cite{langer1969statistical, berera2019formulating}.} of the critical droplet (Sec.~\ref{subsec:prefactor}) yields \cite{langer1969statistical, berera2019formulating}
\begin{equation}
    \sigma = -\frac{\gamma}{2m} + \sqrt{\left(\frac{\gamma}{2m}\right)^2 + \frac{|\lambda_{-1}|}{m}}, \label{eq:sigma}
\end{equation}
where $\lambda_{-1}$ is the negative eigenvalue of the fluctuation kernel $\mathcal{K}\cb$, and the coefficients $m$ and $\gamma$ characterize the effective inertia and dissipation of the field $\theta(x,t)$. Since these coefficients are not known for the experiment \cite{zenesini2024false}, we assume for definiteness that the system is overdamped \cite{fedichev1999dissipative, jackson2007dark, cockburn2011fluctuating} (i.e., \mbox{$\gamma^2\gg 4m|\lambda_{-1}|$}) and express time in units where \mbox{$\gamma=1$} so that \mbox{$\sigma=|\lambda_{-1}|$}.}

\subsection{Critical-droplet contribution}
\label{subsec:inst}

The exponential contribution~\eqref{eq:Ediff} to the nucleation rate depends on the energy difference~$\Delta E$ between the critical droplet $\theta\cb(x)$ and the metastable state $\theta\ms$. These configurations are stationary points of the energy functional~\eqref{eq:energy}, \mbox{$\delta E/\delta \theta = 0$}, and thus solve the Euler-Lagrange equation 
\begin{equation}
    \frac{d^2}{dx^2}\theta = \frac{\partial V}{\partial \theta} = a \cos \theta - \frac{1}{2}\sin 2\theta + c \sin\theta. \label{eq:instanton_equation}
\end{equation}
Solutions of this equation are known as \textit{instantons} in the field-theory literature. The metastable state is homogeneous, so Eq.~\eqref{eq:instanton_equation} reduces to \mbox{$\partial V/\partial\theta |_{\theta\ms}=0$}, where $\theta\ms$ is the highest-energy local minimum of $V(\theta)$ (indicated by the black dashed line in Fig.~\ref{fig:1}(d), for example). The critical droplet $\theta\cb(x)$ is a localized perturbation of the metastable state, i.e., a non-constant solution to Eq.~\eqref{eq:instanton_equation} with boundary condition \mbox{$\theta\cb(|x|\to\infty) = \theta\ms$}. Figures~\ref{fig:1}(c2-e2) show the critical-droplet profile for several points in phase space [corresponding to the red points in Fig.~\ref{fig:1}(a)].

\edit{
Equation~\eqref{eq:instanton_equation} may be interpreted as the equation of motion for a particle with ``position'' $\theta$ at ``time'' $x$ in the inverted potential $-V[\theta]$. This mechanical analogy provides a direct method to obtain the critical droplet configurations. Multiplying Eq.~\eqref{eq:instanton_equation} by $d\theta\cb/dx$ and integrating over the full range of $x$ gives the integral of motion
\begin{equation}
    \frac{1}{2} \left(\frac{d\theta\cb}{dx}\right)^2 = V[\theta\cb(x)] - V[\theta\ms]. \label{eq:intMotion}
\end{equation}
At the center of the critical droplet we have \mbox{$d\theta\cb(x_0)/dx=0$}, which implies \mbox{$\theta\cb(x_0) = \theta\tp$} by Eq.~\eqref{eq:intMotion}. Here, $\theta\tp$ is the turning point, where \mbox{$V[\theta\tp]=V[\theta\ms]$} [indicated by a red dashed line in Fig.~\ref{fig:1}(d)]. The turning point and vanishing derivative provide the initial conditions at $x_0$ for Eq.~\eqref{eq:instanton_equation}. In the mechanical analogy, the critical droplet profile then describes a particle released from rest at $\theta\tp$. Such a particle takes an infinite amount of ``time'' $x$ to roll down the inverted potential and to asymptotically approach the metastable state~$\theta\ms$, yielding a single localized excursion known as a \textit{bounce} instanton [Fig.~\ref{fig:1}(c2,d2,e2)]. Due to the translation symmetry of Eq.~\eqref{eq:instanton_equation}, the center position~$x_0$ of the critical droplet is arbitrary and we set \mbox{$x_0=0$}. This translational symmetry plays no role in the calculation of the exponential contribution to the nucleation rate, but it will contribute when evaluating the prefactor (cf. Sec.~\ref{subsec:prefactor}).}

\begin{figure}[t]
    \centering
    \includegraphics[width=\linewidth]{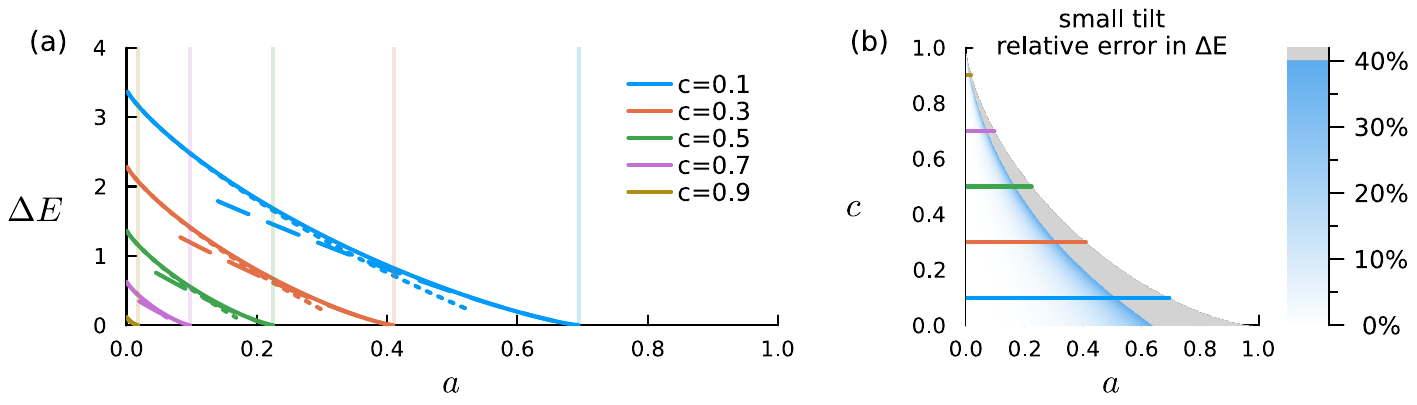}
    \caption{(a) Energy difference $\Delta E$ [Eq.~\eqref{eq:deltaE_full}] between the critical droplet and the metastable state. Dashed and dotted lines show the analytical small-barrier~\eqref{eq:cubicE} and small-tilt~\eqref{eq:E_smalla} approximations, respectively. Transparent vertical lines mark the end points $a_c$ of the metastability region [Eq.~\eqref{eq:ac}]. (b) \edit{Validity region of the small-tilt approximation. Values of the relative error above $40\%$ are colored gray.} Horizontal colored lines correspond to the parameter values in panel (a).}
    \label{fig:2}
\end{figure}

From Eq.~\eqref{eq:intMotion}, we obtain the energy difference between the critical droplet and the metastable state as an integral over the potential barrier, without relying on the instanton profile,
\begin{equation}
    \Delta E = \int_{-\infty}^{\infty} dx\left\{\frac{1}{2}\left(\frac{d\theta\cb}{dx}\right)^2 + V\cb(x) -V\ms\right\} = 2\int_{\theta\tp}^{\theta\ms} d\theta \sqrt{2(V[\theta]-V\ms)}, \label{eq:deltaE_full}
\end{equation}
where \mbox{$V\cb(x)= V[\theta\cb(x)]$} and \mbox{$V\ms= V[\theta\ms]$} (the prefactor $2$ arises because we split the integration from infinity to the turning point and back). Figure~\ref{fig:2}(a) shows the energy difference $\Delta E$ as a function of $a$, obtained numerically for several values of $c$. Intuitively, we expect smaller potential barriers to result in higher nucleation rates, i.e., lower values of $\Delta E$. This is observed in the figure: Near the spinodal line [\mbox{$a\to a_c$}, Eq.~\eqref{eq:ac}], the potential barrier and the critical droplet are shallow [Fig.~\ref{fig:1}(c)]; thus, the value of $\Delta E$ is small. Near the coexistence line [\mbox{$a\to0$}], the height of the potential barrier increases [Fig.~\ref{fig:1}(e)] and so does $\Delta E$.

Close to the spinodal line, an analytical approximation to $\Delta E$ may be found by a cubic-in-$\theta$ expansion of the potential around its inflection point, which gives~\cite{zenesini2024false}
\begin{equation}
    \label{eq:cubicE}
    \Delta E \; \stackrel{a\to a_c}{=} \; \frac{16}{5} 6^{1/4} \, a_c^{-1/4} \, (a_c-a)^{5/4} \, c^{1/6} .
\end{equation}
This small-barrier limit is used in the analysis of Refs.~\cite{zenesini2024false,cominotti2025observation}.  Equation~\eqref{eq:cubicE} is shown as dashed lines in Fig.~\ref{fig:2}(a), and it provides an accurate description of $\Delta E$ near the spinodal line. We do not use this limit in the following as the saddle-point expansion~\eqref{eq:Z} applies for \mbox{$\beta \Delta E\gg 1$}, which excludes the limit \mbox{$a\to a_c$} where  $\Delta E$ approaches zero (cf.~Eq.~\eqref{eq:cubicE} or Fig.~\ref{fig:2}).

Instead, we focus on the limit of small tilts \mbox{$a\to 0$} (i.e., close to the coexistence line), where the minima of $V[\theta]$ are almost degenerate. Unlike for the small-barrier limit, here there is a clear distinction between the background metastable phase and the interior of the critical droplet, which is nearly in the stable phase [cf.~Fig.~\ref{fig:1}(e2)]. The energy in this limit, which we derive in the next subsection, is
\begin{equation}
    \Delta E \; \stackrel{a\to 0}{=} \; 4\left(\sqrt{1-c^2} - c\arccos(c)\right) - 2a \ln\left[\frac{4 e (1-c^2)^{3/2}}{a}\right], \label{eq:E_smalla}
\end{equation}
shown as dotted lines in Fig.~\ref{fig:2}(a). The small-tilt limit is also known as the thin-wall approximation in the literature [in the sense of a thin domain wall, not a small potential barrier, cf. Fig.~\ref{fig:1}(e)]. Figure~\ref{fig:2}(b) shows the relative error of Eq.~\eqref{eq:E_smalla} compared to the exact solution. As is apparent in Fig.~\ref{fig:2}, the small-tilt approximation is accurate over a large fraction of parameter space\edit{: approximately~$55\%$ of all allowed parameter values $a,c$ have a relative error below~$10\%$. For fixed $c$, the approximation covers the range \mbox{$a\lesssim 0.6a_c$}, which overlaps with} the region \mbox{$\beta \Delta E \gg 1$} over which the saddle-point approximation~\eqref{eq:Z} is expected to apply for typical temperatures.

\subsubsection{The small-tilt limit}
\label{sec:small_tilt}

Before discussing fluctuation corrections, we analyze the domain-wall structure of the critical droplet at small potential tilts \mbox{$a\approx0$} in more detail. We will use these results in Sec.~\ref{sec:Gy} to obtain a closed-form expression for the fluctuation determinant, and in Sec.~\ref{sec:Kramers} to provide a simplified description of nucleation in terms of the separation between domain walls.

Along the phase coexistence line~\mbox{$a=0$}, we have \mbox{$\theta\ms = \arccos c$}, and the integral of motion~\eqref{eq:intMotion} reads
\begin{equation}
    \frac{1}{2}\left(\frac{d\theta}{dx}\right)^2 = \frac{1}{2} \left(\cos \theta- c\right)^2,
    \label{eq:kinkInt}
\end{equation}
which upon integration yields
\begin{equation}
    \theta^{\pm}(x-x_0) = \pm 2 \arctan\left(\sqrt{\frac{1-c}{1+c}}\tanh \left[\frac{\sqrt{1-c^2}}{2} (x-x_0)\right]\right). \label{eq:kink}
\end{equation}
This profile describes a single domain wall, or \textit{kink} instanton, which interpolates between the two minima of the potential at $\pm\arccos(c)$ over a region of size \mbox{$(1-c^2)^{-1/2}$} centered at $x_0$. We thus consider the following kink-antikink ansatz for the critical droplet
\begin{equation}
    \theta(x; R) = \theta^+\left(x-\frac{R}{2}\right) + \theta^-\left(x+\frac{R}{2}\right) + \arccos(c), \label{eq:kink_ansatz}
\end{equation}
where the parameter $R$ is the separation between domain walls, and the last term ensures that the ansatz approaches the metastable state as \mbox{$|x|\to\infty$}.

\begin{figure}[t]
    \centering
    \includegraphics[width=\linewidth]{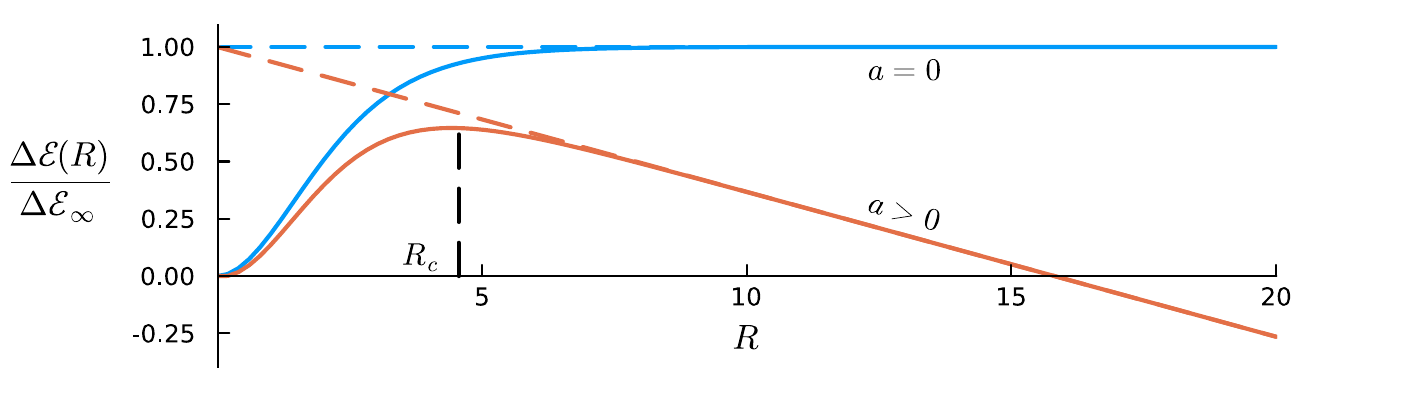}
    \caption{Energy of a kink-antikink pair \eqref{eq:kink_ansatz} as a function of their separation $R$ for \mbox{$c=0.5$}, \mbox{$a=0$} (blue line) and \mbox{$a=0.05$} (orange line). Dashed lines show the respective asymptotic behaviors $\Delta\mathcal{E}_\infty$ and \mbox{$\Delta\mathcal{E}_\infty - \delta V \,R$}. For \mbox{$a>0$}, the critical droplet size $R_c$ (black line) maximizes the energy~\eqref{eq:E(R)_asy}.}
    \label{fig:3}
\end{figure}

The blue curve in Fig.~\ref{fig:3} shows the energy of the ansatz~\eqref{eq:kink_ansatz},
\begin{equation}
    \label{eq:ER}
    \Delta \mathcal{E}(R) = E[\theta(x;R)] - E[\arccos(c)],
\end{equation}
as a function of the domain size $R$ for \mbox{$a=0$}. The minimum \mbox{$\Delta \mathcal{E} = 0$} at \mbox{$R=0$} corresponds to the homogeneous metastable configuration \mbox{$\theta(x;0) = \arccos(c)$}. If the separation between domain walls is larger than their width, \mbox{$R\gg(1-c^2)^{-1/2}$}, expanding the ansatz~\eqref{eq:kink_ansatz} as a single kink plus an exponential tail, \mbox{$\theta(x; R) \approx \theta^+(x-R/2) + 2\sqrt{1-c^2} \,e^{-\sqrt{1-c^2}\left(x+R/2\right)}$}, allows for a perturbative calculation of the energy:
\begin{align}
    \label{eq:asyE}
    \Delta \mathcal{E}(R)|_{a=0} & \stackrel{R\to \infty}{=} \Delta \mathcal{E}_\infty + 4\sqrt{1-c^2}\int_0^\infty dx \, e^{-\sqrt{1-c^2}\left(x+\frac{R}{2}\right)} \; \frac{\delta \mathcal{E}}{\delta \theta}\Biggr|_{\theta^+(x-R/2)} \nonumber \\
    &= \Delta \mathcal{E}_\infty - 8 \left(1-c^2\right)^{3/2} e^{-\sqrt{1-c^2} \,R}.
\end{align}
Here, we use the \mbox{$x\to-x$} symmetry of Eq.~\eqref{eq:kink_ansatz} to integrate over the \mbox{$x>0$} region only, and $\Delta \mathcal{E}_\infty$ is the energy of two isolated kink instantons, 
\begin{equation} \label{DEinfty}
    \Delta \mathcal{E}_\infty = \lim_{R\to\infty} \Delta \mathcal{E}(R)|_{a=0} = 4 \left(\sqrt{1-c^2}-c \arccos c\right).
\end{equation}
This limit is indicated by the dashed blue line in Fig.~\ref{fig:3}, with the first correction to this asymptotic value at large $R$ given by Eq.~\eqref{eq:asyE}.

A small nonzero $a$ lifts the degeneracy of the minima of $V[\theta]$, with an energy difference between the stable and metastable states of \mbox{$\delta V = 2 a \sqrt{1-c^2} + \mathcal{O}(a^2)$}\footnote{The minima of $V[\theta]$ also shift their location, but this only gives a negligible $\mathcal{O}(a^2)$ correction to the energy.}. As a consequence, the energetic cost for creating a kink-antikink pair can now be overcome if they are separated by a large enough region of the energetically favorable stable phase, as discussed in the introduction. This energy competition is illustrated by the orange line in Fig.~\ref{fig:3}, which shows $\Delta \mathcal{E}(R)$ computed using the kink-antikink ansatz~\eqref{eq:ER} for \mbox{$a=0.05$}. Starting from the metastable state at \mbox{$R=0$}, the energy increases for small $R$ but then decreases asymptotically as \mbox{$R \delta V = 2Ra \sqrt{1-c^2}$}, so that large domains become energetically favorable.

\begin{figure}[b]
    \centering
    \includegraphics[width=\linewidth]{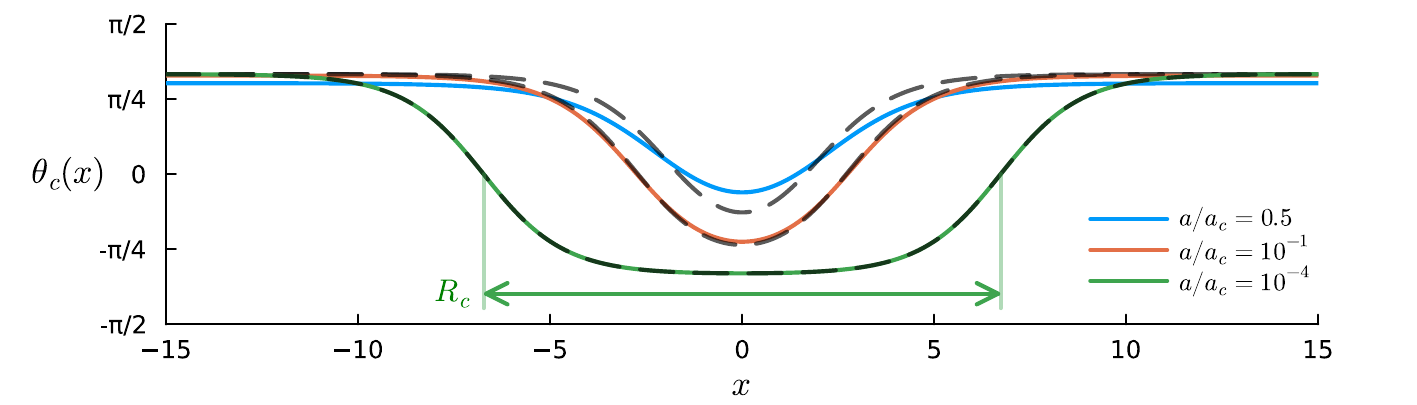}
    \caption{Critical droplet as \mbox{$a\to0$} (solid lines), obtained by numerically solving Eq.~\eqref{eq:instanton_equation} for \mbox{$c=0.5$.} Dashed lines show the corresponding kink-antikink pairs [Eq.~\eqref{eq:kink_ansatz}] separated by a distance \mbox{$R=R_c$} [Eq.~\eqref{eq:Rcrit}].}
    \label{fig:4}
\end{figure}

Combining the asymptotic expansion~\eqref{eq:asyE} with the tilt contribution $-R \delta V$, we obtain the energy of the ansatz~\eqref{eq:kink_ansatz} at small $a$ and large $R$
\begin{equation}
    \Delta \mathcal{E}(R) \sim \Delta \mathcal{E}_\infty - 8 (1-c^2)^{3/2} e^{-\sqrt{1-c^2} \,R} - 2Ra \sqrt{1-c^2} . \label{eq:E(R)_asy}
\end{equation}
The maximum of the resulting energy barrier determines the critical droplet size
\begin{equation}
    R_c = \frac{1}{\sqrt{1-c^2}}\ln\Biggl[\frac{4 (1-c^2)^{3/2}}{a}\Biggr], \label{eq:Rcrit}
\end{equation}
explicitly expressed as a multiple of the domain-wall width $1/\sqrt{1-c^2}$. Figure~\ref{fig:4} shows that the ansatz~\eqref{eq:kink_ansatz} with \mbox{$R=R_c$} (dashed lines) agrees well with the exact critical-droplet profiles obtained from a numerical solution of Eq.~\eqref{eq:instanton_equation} (solid lines). Remarkably, the kink-antikink ansatz works well even beyond the strict small-tilt limit, i.e., when $R_c$ is comparable to the width of the domain walls (orange line).

\subsection{Fluctuation contribution}
\label{subsec:prefactor}

We now evaluate the equilibrium prefactor to the nucleation rate~\eqref{eq:Zratio}, which arises from small fluctuations around the critical droplet and the metastable state. This requires computing a functional Gaussian integral~\eqref{eq:Z0_temp1} and the resulting fluctuation determinant~\eqref{eq:D}. In the quantum-field-theory literature, this contribution is known as a \textit{one-loop correction} due to its diagrammatic representation~\cite{zinn2021quantum}.

To evaluate the functional integral~\eqref{eq:Z0_temp1}, we decompose $\delta\theta$ into eigenfunctions $q_n$ of the fluctuation operator $\mathcal{K}_{\rm sp}$~\eqref{eq:K},
\begin{equation}
    \delta\theta(x) = \sum_n c_n \,q_n(x), \label{eq:expansiontheta}
\end{equation}
where $q_n(x)$ solves
\begin{equation}
    \mathcal{K}_{\rm sp} \, q_n(x) = \Biggl(-\frac{d^2}{dx^2} + \frac{\partial^2 V[\theta]}{\partial \theta^2} \biggr|_{\theta_{\rm sp}(x)}\Biggr) \edit{q_n(x)} = \lambda_n q_n(x).
\end{equation}
In the thermodynamic limit \mbox{$L\to\infty$}, bulk properties like the nucleation rate are independent of the choice of boundary conditions; we consider Dirichlet boundary conditions, \mbox{$q(\pm L/2)=0$}, in the following. The expansion~\eqref{eq:expansiontheta} decouples the functional integral~\eqref{eq:Z0_temp1} into a product of eigenvalues\footnote{\edit{Here, we use the normalized measure \mbox{$\mathcal{D}[\delta\theta] = \prod_n dc_n/\sqrt{2\pi \beta^{-1}}$}. The nucleation rate depends on the ratio of partition functions~\eqref{eq:Zratio}, so factors in the integral measure do not affect the final result \cite{coleman1985selected}.}}
\begin{equation}
    \mathcal{Z}_{\rm sp} = e^{-\beta E_{\rm sp}} \int \prod_n \frac{dc_n}{\sqrt{2\pi \beta^{-1}}}\; \exp\left(-\frac{\beta}{2} \lambda_n c_n^2\right) = e^{-\beta E_{\rm sp}} \prod_n \lambda_n^{-1/2} = \frac{e^{-\beta E_{\rm sp}}}{\sqrt{\det \mathcal{K}_{\rm sp}}}.
    \label{eq:asEigs}
\end{equation}
This expression is complete for fluctuations around the metastable state, as the eigenvalues of $\mathcal{K}\ms$ are all positive. This positivity is a result of the metastable state being a minimum of the energy functional: Small deviations from this configuration always increase the energy. For~$\mathcal{K}\cb$, however, although most fluctuations around the critical droplet increase the energy, there are two important exceptions that must be treated separately: The unstable mode and the zero mode of the critical droplet.

The first exception arises because the critical droplet is a saddle point of the energy, meaning $\mathcal{K}\cb$ has a negative eigenvalue \mbox{$\lambda_{-1}<0$}. The associated \textit{unstable mode} describes a fluctuation that increases the amount of the stable phase in the critical droplet, decreasing its energy and precipitating nucleation. The Gaussian integral of $\lambda_{-1}$ is formally divergent, which signals that the expansion is not around a true equilibrium state. The determinant~\eqref{eq:asEigs} is then defined via analytic continuation\footnote{\edit{While analytic continuation is necessary in the functional-integral approach used here, it is  not required in other derivations of the nucleation rate \cite{berera2019formulating, langer1969statistical}.}}, which gives an imaginary contribution of this eigenvalue \edit{$i/(2\sqrt{|\lambda_{-1}|})$}~\cite{callan1977fate}.

\edit{The unstable mode also determines the dynamical prefactor $\sigma$. Substituting the time-dependent perturbation to the critical droplet \mbox{$\theta(x,t) = \theta\cb(x) + \delta\theta(x,t)$} into Eq.~\eqref{eq:langer_dynamics} yields
\begin{equation}
    m\frac{\partial^2}{\partial t^2} \delta\theta + \gamma\frac{\partial}{\partial t} \delta\theta = - \left(-\frac{d^2}{dx^2} + V''\cb\right)\delta\theta = -\mathcal{K}\cb\delta\theta,
\end{equation} 
where we neglect the noise term for the stability analysis. Expanding the perturbation in eigenmodes \eqref{eq:expansiontheta} and projecting onto the $n$-th mode gives
\begin{equation}
    m\ddot{c}_n(t) + \gamma\dot{c}_n(t) = - \lambda_n c_n(t),
\end{equation} 
for the evolution of the $n$-th mode amplitude. The dynamical prefactor $\sigma$ is identified as the unstable-mode growth rate, \mbox{$c_{-1} \sim e^{\sigma t}$}, which yields Eq.~\eqref{eq:sigma}}.

The second exception is due to translation symmetry, as it is possible for a fluctuation to simply shift the critical droplet without changing its energy. The corresponding eigenmode of $\mathcal{K}\cb$ with \mbox{$\lambda_0=0$}, called the~\textit{zero mode}, is
\begin{equation}
    \label{eq:q_zero_mode}
    q_0(x) = \frac{1}{\sqrt{\Delta E}} \; \frac{\partial\theta\cb}{\partial x} ,
\end{equation}
where $\partial \theta\cb/\partial x$ is the change in $\theta\cb$ due to an infinitesimal translation, and $1/\sqrt{\Delta E}$ is a normalization factor. Just as for the unstable mode, the integral over the zero mode is formally divergent. To evaluate it, we use an alternative parametrization of the fluctuations around~$\theta\cb$~\cite{kleinert2006path},
\begin{equation} \label{eq:collectivecoordinate}
    \theta(x) = \theta\cb(x) + \sum_{n} c_n \, q_n(x) = \theta\cb(x-x_0) +\sum_{n\neq 0} c_n \, q_n(x),
\end{equation}
where we exclude the zero mode from the expansion and introduce instead the position of the critical droplet $x_0$ as a collective coordinate. Performing the change of coordinates \mbox{$dc_0\to dx_0$} using Eq.~\eqref{eq:collectivecoordinate} in the integration measure of Eq.~\eqref{eq:asEigs} results in
\begin{equation}
    \sqrt{\frac{\beta}{2\pi}} \int_{-\infty}^\infty dc_0 = \sqrt{\frac{\beta}{2\pi}} \int_{-L/2}^{L/2} dx_0 \, \left| \frac{1}{q_0}\frac{\partial\theta\cb}{\partial x}\right|_{x=x_0}
    = L \left(\frac{\beta \Delta E}{2\pi}\right)^{1/2}.
    \label{eq:zero_mode}
\end{equation}
This contribution is proportional to the system size $L$, as is expected for an extensive quantity like the nucleation rate.

We have not yet specified how to evaluate the fluctuation determinant $D(a,c)$  [Eq.~\eqref{eq:D}]. We now proceed to compute it over the full parameter space, including a closed-form expression in the small-tilt regime.

\subsubsection{Gel'fand-Yaglom evaluation of the fluctuation determinant}
\label{sec:Gy}

A direct evaluation of the fluctuation determinant $D(a,c)$ would involve obtaining the spectra of the fluctuation operators $\mathcal{K}\ms$ and $\mathcal{K}\cb$ and assembling the divergent products of eigenvalues into a finite result. Usually, this has to be done numerically, so additional effort is needed to determine the asymptotic behavior of the eigenvalues and to deal with a continuum of scattering states. Here, we use a different approach, originally due to Gel'fand and Yaglom \cite{gel1960integration}, which simplifies the evaluation of the fluctuation determinant to a single initial value problem. To benchmark our results, we also derive them by direct diagonalization in Appendix~\ref{app:direct_diagonalization}.

The central idea of the Gel'fand-Yaglom approach is to evaluate the functional determinant through a characteristic function, which avoids a full evaluation of the fluctuation spectrum. To briefly introduce this idea, consider a matrix with eigenvalues $\lambda_i$ and define the characteristic polynomial \mbox{$\mathcal{P}(\lambda) = \prod_i (\lambda_i - \lambda)$}. The roots of $\mathcal{P}(\lambda)$ coincide in location and multiplicity with the eigenvalues, so its constant term $\mathcal{P}(0)$ is proportional to the determinant of the matrix. A similar result holds for functional determinants: For an analytic function $\mathcal{P}(\lambda)$ whose zeros correspond in location and multiplicity to the eigenvalues of a differential operator, $\mathcal{P}(0)$ is proportional to the functional determinant of that operator \cite{kirsten2004functional, dunne2008functional, ekstedt2023bubbledet, garcia2024fluctuation}. $\mathcal{P}(\lambda)$ is known as the Evans function in the mathematical literature~\cite{kapitula2013spectral}.

To construct the function $\mathcal{P}_{\rm sp}(\lambda)$ for the differential operator $\mathcal{K}_{\rm sp}$, let \mbox{$\mathcal{P}_{\rm sp}(\lambda) = u_\lambda(L/2)$}, where $u_\lambda(x)$ solves
\begin{equation}
    \mathcal{K}_{\rm sp}\,u_\lambda(x) = \Biggl(-\frac{d^2}{dx^2} + \frac{\partial^2 V}{\partial \theta^2} \biggr|_{\theta_{\rm sp}(x)}\Biggr) \, u_\lambda(x) = \lambda u_\lambda(x) \label{eq:GY_ODE}
\end{equation}
with initial conditions \mbox{$u_\lambda|_{x=-L/2}=0$} and \mbox{$du_\lambda/dx|_{x=-L/2}=1$} at the left boundary. Whenever \mbox{$\mathcal{P}_{\rm sp}(\lambda)=u_\lambda(L/2)=0$}, Dirichlet boundary conditions are satisfied at both endpoints. Thus, $\mathcal{P}_{\rm sp}(\lambda)$ is a smooth function of $\lambda$ such that \mbox{$\mathcal{P}_{\rm sp}(\lambda) = 0$} if and only if $\lambda$ is an eigenvalue of~$\mathcal{K}_{\rm sp}$; by our previous discussion, the fluctuation determinant in Eq.~\eqref{eq:D} is then given by
\begin{equation}
    D(a,c) = \lim_{\lambda\to 0} \frac{1}{\lambda}\frac{\mathcal{P}\cb(\lambda)}{\mathcal{P}\ms(0)}, \label{eq:GY_det}
\end{equation}
where the limiting procedure removes the contribution of the zero mode. We stress that evaluating the expression on the right-hand side involves only solving Eq.~\eqref{eq:GY_ODE} for vanishingly small values of $\lambda$, which is much simpler than finding every eigenvalue and evaluating their product. Appendix~\ref{app:numeric_GY} provides details on the numerical evaluation of Eq.~\eqref{eq:GY_det}.

Equation~\eqref{eq:GY_det} extends to arbitrary potentials in the energy functional as well as to higher dimensions. In the particular case of 1D systems with translation symmetry, it may be further simplified to~\cite{kleinert2006path}
\begin{equation}
    \edit{D(a,c) = \frac{\Delta E}{2k^3\Theta^2}}, \label{eq:magic}
\end{equation}
where $\Theta$ and $k$ are, respectively, the amplitude and the spatial decay rate of the critical droplet to its asymptotic metastable value, \mbox{$\theta\cb(x) \sim \theta\ms - \Theta e^{-k |x|}$}. In the small-tilt limit, our results for the energy and shape of the critical droplet [Eqs.~\eqref{eq:E_smalla} and~\eqref{eq:kink_ansatz}] yield
\begin{equation}
    D(a, c) \; \stackrel{a\to 0}{=} \; \frac{\sqrt{1-c^2}-c \arccos c}{8 \left(1-c^2\right)^4} a. \label{eq:smallTiltD}
\end{equation}

\begin{figure}[t]
    \centering
    \includegraphics[width=\linewidth]{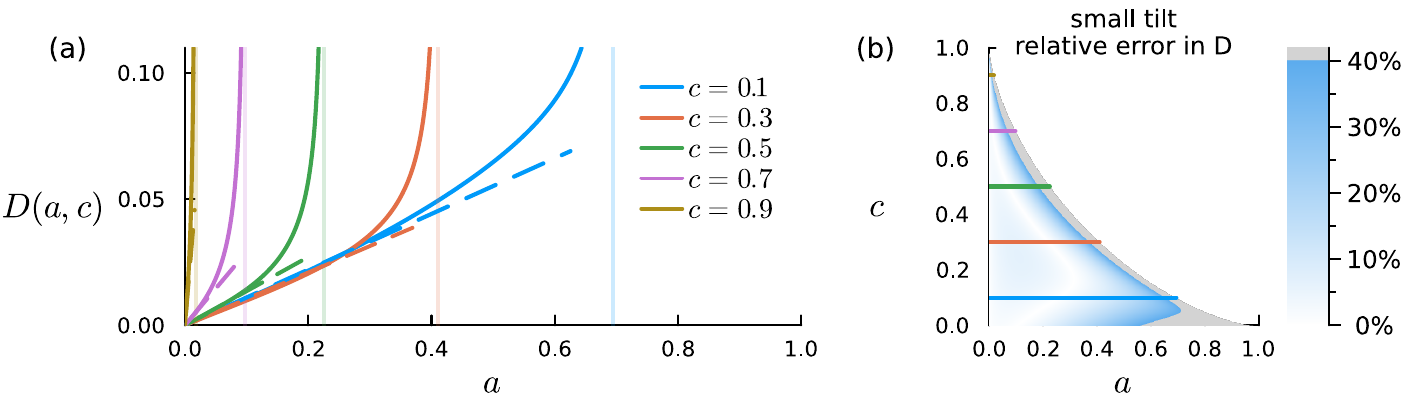}
    \caption{(a) Fluctuation determinant $D(a,c)$ evaluated using the Gel'fand-Yaglom method~\eqref{eq:GY_det}. Vertical lines mark the boundary of the metastability region~\eqref{eq:ac}. Dashed lines show the small-tilt approximation~\eqref{eq:smallTiltD}. (b)~\edit{Error of the small-tilt approximation relative to the exact result. Relative errors above $40\%$ are colored gray.}
    }
    \label{fig:5} 
\end{figure}

Figure~\ref{fig:5}(a) shows the fluctuation determinant $D(a,c)$ over the full parameter space obtained by numerically solving Eq.~\eqref{eq:GY_ODE}. It is immediately apparent that the fluctuation determinant has a significant dependence on the parameters $a$ and $c$. In particular, simplifying treatments that replace the determinant by a (phenomenological) constant do not apply. The small-tilt result~\eqref{eq:smallTiltD} closely follows the exact result for \mbox{$a\lesssim a_c/2$} [Fig.~\ref{fig:5}(a), dashed lines]. In this regime, the kink-antikink form of the critical droplet [Eq.~\eqref{eq:kink_ansatz}] intuitively explains why the fluctuation determinant vanishes as \mbox{$a\to 0$}: The unstable mode of the critical droplet expands the stable-phase domain by increasing the separation between domain walls. At~\mbox{$a=0$}, however, the domain walls are themselves kink instantons with their own translation zero modes. This indicates that, in the small-tilt limit, the unstable mode of the critical droplet becomes an additional zero mode of the kink instantons, explaining the limiting behavior \mbox{$\lim_{a\to 0} D(a,c) = 0$}. Figure~\ref{fig:5}(b) shows a comparison of the relative error between the full result~\eqref{eq:GY_det} for the fluctuation determinant and the small-tilt limit~\eqref{eq:smallTiltD}. Consistent with Fig.~\ref{fig:2}(b), the small-tilt approximation \edit{is below $10\%$ relative error for \mbox{$a\lesssim 0.6a_c$}, which covers approximately $58\%$ of all allowed values of the parameters $a,c$.}

We note that the results for the fluctuation determinant are independent of the parametrization of the energy functional~\eqref{eq:energy}. For example, expressing it directly in terms of the imbalance \mbox{$Z(x) = (n_{\uparrow} - n_\downarrow)/n$} instead of $\theta(x)$ gives identical results. Such a transformation, however, introduces a technical complication in the form of a field-dependent stiffness in the energy functional, which leads to a divergent fluctuation contribution that cancels with a corresponding divergence in the path-integral measure. We discuss these details in Appendix~\ref{app:fdkt}.

\subsection{The nucleation rate}
\label{subsec:result}

\begin{figure}[t]
    \centering
    \includegraphics[width=\linewidth]{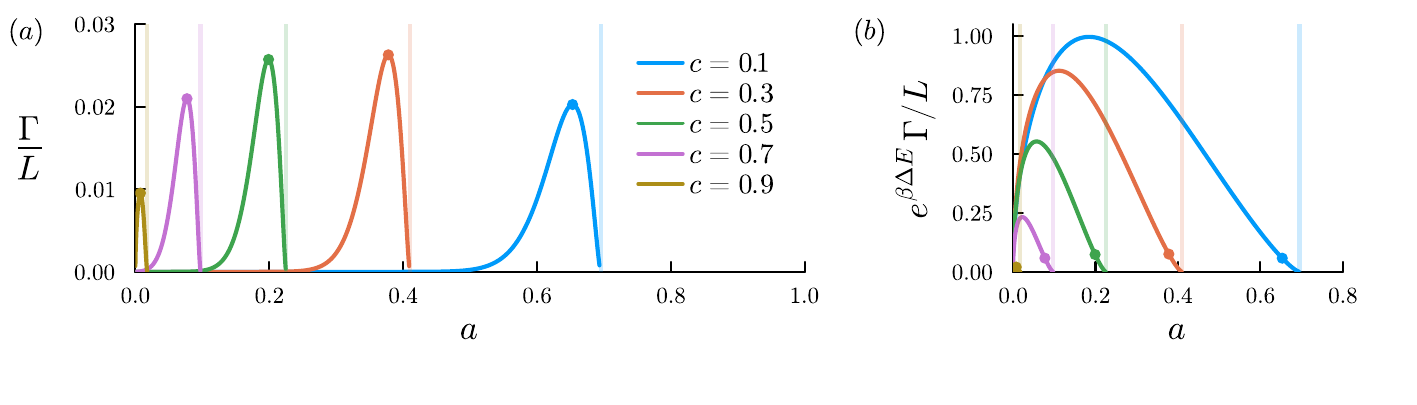}
    \caption{(a) Nucleation rate $\Gamma/L$ per unit length [Eq.~\eqref{eq:l_rate}] evaluated for \mbox{$\beta=15$}. Vertical lines mark the boundary of the metastability region~\eqref{eq:ac}. The peaks of the nucleation rate, marked by dots at \mbox{$\Delta E \sim \beta^{-1}$}, bound the region of applicability of our results. To the left of a peak we have \mbox{$\beta\Delta E> 1$}, which is required for the saddle-point evaluation of Eq.~\eqref{eq:Z}. In this region the nucleation rate decreases with increasing barrier height (decreasing $a$). To the right of a peak, the saddle-point condition fails and we have an unphysical decay of the rate, even as the potential barrier vanishes. \edit{(b) Nucleation rate prefactor. Colored dots correspond to those in panel (a).}} 
    \label{fig:6}
\end{figure}

\edit{Figure~\ref{fig:6} shows our combined result for the nucleation rate, where Fig.~\ref{fig:6}(a) shows the nucleation rate $\Gamma/L$ per unit length [Eq.~\eqref{eq:l_rate}], and Fig.~\ref{fig:6}(b) shows the prefactor, i.e., the nucleation rate $e^{\beta \Delta E}\Gamma/L$ excluding the Arrhenius term. We use an inverse temperature \mbox{$\beta=15$} that is representative of typical experimental values in~\cite{zenesini2024false}\footnote{Note our choice of dimensionless energy (cf.~footnote \ref{footnote:1}) differs from that of \cite{zenesini2024false} by a factor of $2\sqrt{2}$.}. We obtain the overdamped dynamical factor \mbox{$\sigma=|\lambda_{-1}|$} by numerically solving the eigenvalue equation~\eqref{eq:GY_ODE}. The rate prefactor is often neglected or approximated in simplified treatments, but Fig.~\ref{fig:6}(b) explicitly shows that the prefactor depends non-trivially on the control parameters $a$, $c$, and cannot be approximated by a constant value.}

\edit{
Near the phase coexistence line at \mbox{$a=0$}, the unstable mode has the asymptotic form\footnote{To this order in $a$, the result follows from the variational method applied to the kink-antikink ansatz.}
\begin{equation}
    \lambda_{-1} \; \stackrel{a\to 0}{=} \; -a\frac{2(1-c^2)}{\sqrt{1-c^2} - c \arccos c},
\end{equation}
and thus the nucleation rate vanishes as
\begin{equation}
    \frac{\Gamma}{L} \; \stackrel{a\to 0}{=} \; \frac{4}{\pi^{3/2}} \frac{(1-c^2)^3}{\sqrt{1-c^2} - c \arccos c} \sqrt{\beta a} \; e^{-\beta \Delta E}.
\end{equation}
In this regime, the potential~\eqref{eq:V} becomes symmetric and it is possible for either phase to nucleate in the other. Thus, the system reaches an equilibrium state characterized by the formation and annihilation of kink-antikink pairs, known as an instanton gas \cite{habib2000dynamics, bray1993theory, altland2010condensed}. Increasing the tilt from \mbox{$a=0$} exponentially increases the rate to experimentally observable values.}

\edit{Intuitively, we expect the rate of nucleation to keep increasing as the potential barrier decreases in energy, i.e., as the tilt $a$ increases. However, within our model increasing the tilt to values close to the spinodal line causes the nucleation rate to decrease instead [Fig.~\ref{fig:6}(a)]. This should not be considered a physical prediction of the theory; rather, it indicates its limit of applicability. Indeed, the rate~\eqref{eq:l_rate} depends on temperature as \mbox{$\Gamma\sim \beta^{1/2}e^{-\beta\Delta E}$}, which has a maximum with respect to $\beta$ at \mbox{$\beta \Delta E = 1/2$}. However, the saddle-point evaluation of Eq.~\eqref{eq:Z} \edit{is formally an expansion in the parameter} \mbox{$\beta \Delta E \gg 1$} \cite{ekstedt2022bubble}. Thus, the maxima of Fig.~\ref{fig:6}(a) signal a point beyond which the thermal nucleation theory breaks down. This reinforces our discussion at the end of Sec.~\ref{subsec:inst}, where we focus on the small-tilt instead of the small-barrier regime for our nucleation theory.}

\section{Domain formation and growth as an effective escape problem}
\label{sec:Kramers}

In the previous section we focused on the stochastic nucleation rate of critical droplets. Once a droplet grows past its critical size, however, its evolution is essentially deterministic and fixed by the universality class of the system \cite{hohenberg1977theory, bray1993theory}. This implies two characteristic time scales: the rate at which critical droplets form (i.e., nucleate) and the rate at which they grow once formed. While a full description of domain dynamics is beyond the scope of this paper, distinguishing the two time scales in experiments is important for a precise measurement of the nucleation rate. Thus, in this section we develop an effective theory of domain growth based on our analytical results for the small-tilt limit.

We have shown in Sec.~\ref{sec:small_tilt} that the individual domain walls of the critical droplet in the small-tilt limit are described by kink instantons~\eqref{eq:kink}. The fluctuation spectrum of an isolated kink instanton contains a translation zero mode and modes with positive eigenvalues, but no negative-eigenvalue modes. This implies that domain walls are robust against deformations of their shape but have a low energy cost for rigid translations, suggesting that the main effect of thermal noise on a nucleated domain will be the diffusion of its domain walls. This motion is paired with energy-minimizing deterministic dynamics, leading to the growth of large domains and the reduction of small ones. We thus formulate the nucleation problem in terms of a Langevin equation for the separation $R$ between domain walls, effectively providing a link between \edit{the classical reaction-coordinate approach to nucleation and the field-theoretical formulation that leads to Eq.~\eqref{eq:l_rate}}. The idea of projecting the full dynamics of a field onto a single collective coordinate has been applied to other instanton models, such as the sine-Gordon chain \cite{marchesoni1987brownian}, the $\phi^4$ model \cite{habib2000dynamics}, and bright solitons in superfluids \cite{efimkin2016non}.

\subsection{A homogeneous fluid}
\label{sec:homogeneous_fluid}

\edit{
We adopt a phenomenological approach and model a nucleating droplet as a pair of domain walls of fixed shapes separated by a distance $R$. This separation is subject to thermal fluctuations and deterministic evolution as dictated by the potential $\Delta \mathcal{E}(R)$ [Eq.~\eqref{eq:ER}, Fig.~\ref{fig:3}]. In the overdamped limit, the collective coordinate $R$ evolves according to
\begin{equation}
    \label{eq:kramer_R} 
    \frac{dR}{dt} = -\frac{1}{\gamma_K} \frac{d}{dR} \Delta \mathcal{E}(R) + \; \xi(t), 
\end{equation}
where $\gamma_K$ is the effective dissipation coefficient experienced by the domain walls, and $\xi(t)$ is a Gaussian white noise with variance \mbox{$\langle \xi(t)\xi(t')\rangle=2 \delta(t-t')/(\beta\gamma_K)$} set by the fluctuation-dissipation theorem~\cite{zwanzig2001nonequilibrium}. This effective description simplifies the infinite number of degrees of freedom of the full field-theoretical nucleation problem to a single one, the size $R$ of the stable-phase domain. The dynamics \eqref{eq:kramer_R} define a Kramers escape problem, where the escape and expansion of $R$ correspond to the nucleation and relaxation of the full field $\theta(x,t)$ to the stable phase.}

\begin{figure}[t]
    \centering
    \includegraphics[width=\linewidth]{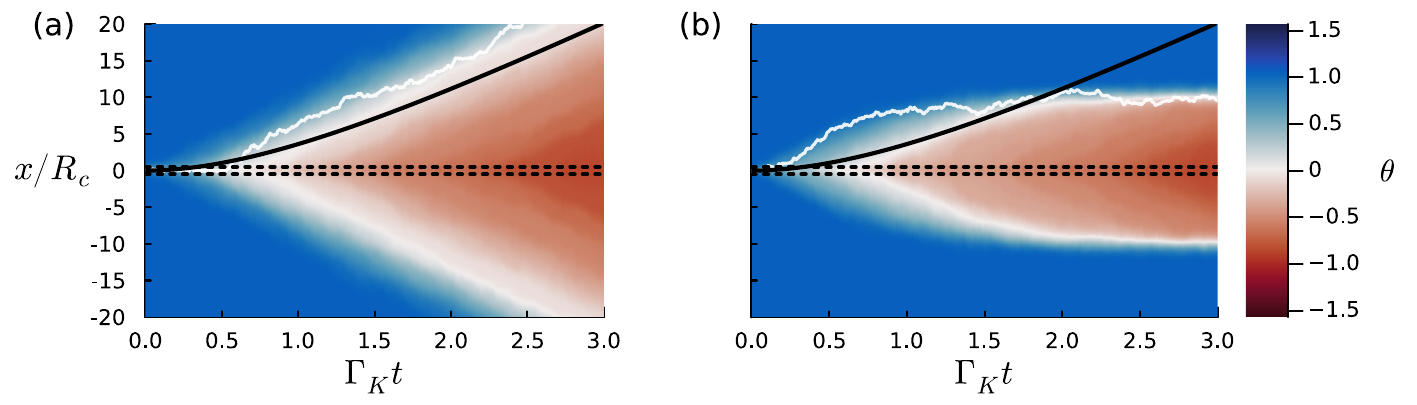}
    \caption{Intensity plot of the average polarization $\langle\theta(x; R(t))\rangle$ as obtained from $100$ simulations of Eq.~\eqref{eq:kramer_R} with the small-tilt ansatz~\eqref{eq:ER} for \mbox{$a=0.1$}, \mbox{$c=0.5$}, \edit{\mbox{$\beta=5$}}, and \mbox{$\gamma_K=1$}. Panel (a) shows a homogeneous system~\eqref{eq:kramer_R}, while in panel (b) we use the position-dependent energy~\eqref{eq:density_E} with a harmonic trap of size \mbox{$R_{\rm TF}=40R_c$}. White lines show individual realizations of the trajectory, solid black lines show the theoretical prediction~\eqref{eq:ensR} for $\langle R(t)\rangle$, and dotted black lines indicate the critical-droplet size $R_c$.}
    \label{fig:7}
\end{figure}

\edit{
A typical trajectory of the stochastic differential equation~\eqref{eq:kramer_R} [white line in Fig.~\ref{fig:7}(a)] fluctuates around \mbox{$R=0$} before crossing the energy barrier at \mbox{$R=R_c$} [cf.~Fig.~\ref{fig:3}] \footnote{For the numerical calculation, we use the \texttt{SOSRA} algorithm in the \texttt{DifferentialEquations.jl} library.}. The crossing rate in the low-temperature limit is given by the Kramers rate~\cite{zwanzig2001nonequilibrium, berera2019formulating}
\begin{align}
    \Gamma_K = \frac{\sigma_K}{2\pi} \times {\sqrt{\frac{\partial_R^2 \, \Delta \mathcal{E}(0)}{|\partial_R^2 \, \Delta \mathcal{E}(R_c)|}}} \; \, e^{-\beta \Delta\mathcal{E}(R_c)}, \label{eq:Kramers_G}
\end{align}
where the overdamped dynamical prefactor is
\begin{equation}
    \sigma_K =
    \frac{|\partial_R^2 \, \Delta \mathcal{E}(R_c)|}{\gamma_K}.
\end{equation}
The Kramers escape} is a memoryless process---i.e., the rate $\Gamma_K$ is constant over time and does not depend on previous history---resulting in an exponential distribution for the escape times
\begin{equation}
\label{eq:nucleatedfraction}
    {\rm Prob}\left(t_{\rm escape} \leq t\right) = 1 - e^{-\Gamma_K t}.
\end{equation}
\edit{
Interestingly, the observed fraction of experiments \cite{zenesini2024false} with nucleated domains is consistently smaller than the memoryless prediction~\eqref{eq:nucleatedfraction}, a discrepancy that could be due to difficulties in identifying the critical bubble in the low-contrast regime [Fig.~\ref{fig:1}(c2)].} Shortly after escape, the deterministic dynamics takes over and $R$ increases with a constant average expansion speed given by
\begin{equation}
    v_\infty = \lim_{R\to\infty} \left(-\frac{1}{\gamma_K}\frac{d}{dR} \Delta \mathcal{E}(R)\right) = \frac{1}{\gamma_K}\delta V = \frac{2a}{\gamma_K}\sqrt{1-c^2},
\end{equation}
where, as before, $\delta V$ is the energy difference between the stable and metastable states of the potential at small tilts.

\edit{Figure~\ref{fig:7}(a) shows the ensemble-averaged polarization field, $\langle \theta(x;R(t))\rangle$, computed from an ensemble of trajectories satisfying the Langevin equation~\eqref{eq:kramer_R}. The red and blue regions denote the stable and metastable phases, respectively. To obtain an analytical approximation, we assume that each trajectory remains at \mbox{$R=0$} until a stochastic escape time, drawn from an exponential distribution with rate $\Gamma_K$, and that $R(t)$ subsequently increases linearly with speed $v_\infty$. Neglecting the finite width of the small-tilt domain walls, the local field takes the metastable value \mbox{$\theta=\theta\ms$} when \mbox{$|x|>R(t)/2$}, and the stable value \mbox{$\theta=\theta_{\rm s}$} otherwise. Averaging over the escape-time distribution then yields
\begin{equation}
    \langle \theta(x;R(t))\rangle = \theta\ms e^{-\Gamma_K\left[t-\frac{2|x|}{v_\infty}\right]_+} + \theta_{\rm s}\left(1-e^{-\Gamma_K\left[t-\frac{2|x|}{v_\infty}\right]_+}\right), \label{eq:1_11_theory} 
\end{equation} 
where \mbox{$[y]_+=\max\{0,y\}$}. Within the same approximation, the boundary between the two phases is given by
\begin{equation}
    \langle R(t)\rangle = \int_0^\infty ds \, \big[v_\infty(t-s)\big]_+ \, \Gamma_K e^{-\Gamma_K s} = v_\infty \biggl(t - \frac{1-e^{-\Gamma_K t}}{\Gamma_K} \biggr),
    \label{eq:ensR}
\end{equation}
which is shown as a continuous black line in Fig.~\ref{fig:7}(a). This equation illustrates the different contributions of the two time scales: the nucleation rate $\Gamma_K$, which depends on the energy difference between the metastable state and critical droplet, and the expansion rate $v_\infty/R_c$, which depends on the energy difference between the metastable and stable phases.}

We note the formal similarity between the escape rate~\eqref{eq:Kramers_G} and the nucleation rate~\eqref{eq:l_rate}. \edit{Separating from the fluctuation determinant $D$ the unstable mode of the critical droplet and the two lowest modes of the metastable state ($\lambda_{m,0}$,$\lambda_{m,1}$), and exponentiating the remaining contribution yields
\begin{equation}
    \Gamma = \frac{\sigma}{2\pi} \sqrt{\frac{\lambda_{m,0}\lambda_{m,1}}{|\lambda_{-1}|}} \; L\left(\frac{\beta \Delta E}{2\pi}\right)^{\frac{1}{2}} \; e^{-\beta \left(\Delta E + \frac{1}{2\beta} \log\tilde D\right)}.
\end{equation}
Here, \mbox{$\tilde D= D \lambda_{m,0}\lambda_{m,1}/|\lambda_{-1}|$}, so that \mbox{$\log\tilde D/(2\beta)$} may be interpreted as a fluctuation correction to the activation barrier~\cite{gould2021effective}. In the thin-wall limit, we expect that the eigenvalues $\lambda_{m,0}$ and $|\lambda_{-1}|$ map to the respective curvature terms in Eq.~\eqref{eq:Kramers_G}, while the remaining metastable eigenvalue gives the additional factor \mbox{$\sqrt{\lambda_{m,1}}\approx \sqrt{1-c^2}$}. However, a complete matching would require relating the normalized field-theory eigenvalues $|\lambda_{-1}|$, $\lambda_{m,0}$ to the curvatures $\partial_R^2\Delta\mathcal E$ appearing in Eq.~\eqref{eq:Kramers_G}, which is beyond the scope of the current work.}

\subsection{A harmonically trapped fluid}

Experiments with cold gases take place inside a trap, which induces a position-dependent density profile neglected in our previous discussion. Assuming the critical droplet size is small compared to the trap length, we obtain the total nucleation rate from a local density approximation
\begin{equation}
    \Gamma = \int dx \, \frac{\Gamma[n(x)]}{L}.
\end{equation}
Using the same assumptions as in the previous section, we give a qualitative prediction for the effect of a position-dependent density on the average profile of growing droplets.

Based on experimental observations \cite{zenesini2024false}, we assume a single droplet nucleates at the center of the trap and expands to form an extended stable domain. Neglecting density gradients, the modified energy functional~\eqref{eq:energy} including a position-dependent local density reads
\begin{equation}
    \Delta \mathcal{E}(R) = \int_{-R_{\rm TF}/2}^{R_{\rm TF}/2} dx\frac{n(x)}{n(0)}\left\{\frac{1}{2}\left(\frac{d\theta(x;R)}{dx}\right)^2 + V[\theta(x;R)] -V\ms\right\}, \label{eq:density_E}
\end{equation}
where we use $n(0)$ in the dimensionless definitions of $a$ and $c$. For harmonic confinement, the density profile in the Thomas-Fermi approximation is a parabola~\cite{pethick2008bose}
\begin{equation}
    \frac{n(x)}{n(0)} = 1-4\frac{x^2}{R_{\rm TF}^2}.
    \label{eq:TF}
\end{equation}
Separating the energy into bulk and domain-wall contributions, we obtain
\begin{equation}
    \Delta \mathcal{E}_{\rm TF}(R) = \Delta \mathcal{E}_\infty \frac{n(R/2)}{n(0)} - \delta V \int_{-R/2}^{R/2} dx\, \frac{n(x)}{n(0)} = \Delta \mathcal{E}_\infty \left(1 - \frac{R^2}{R_{\rm TF}^2}\right) - \delta V \left(R - \frac{R^3}{3 R_{\rm TF}^2}\right).
\end{equation}
Figure~\ref{fig:7}(b) shows the average polarization field over many realizations of the escape problem for the energy~\eqref{eq:density_E} with the density profile~\eqref{eq:TF}. As predicted, when $\langle R\rangle$ is small, its evolution coincides with the homogeneous density case~\eqref{eq:ensR}. As the average size of the domain increases, its \edit{expansion speed} decreases with the distance to the edge of the condensate, consistent with experimental observations~\cite{zenesini2024false}.

\edit{
\subsection{Experimental observables}}

\edit{
Previous experiments \cite{zenesini2024false, cominotti2025observation} determine the nucleation rate by probing the fraction of condensate in the metastable phase \mbox{$F_t = 1 - \langle R\rangle/R_{\rm TF}$}. Equation~\eqref{eq:ensR} predicts the initial evolution \mbox{$F_t \sim 1- t^2$}, in qualitative agreement with the empirical function used by~\cite{zenesini2024false}. However, measuring~$F_t$ does not directly yield the nucleation rate in our model, as both its early-time behavior~\eqref{eq:ensR},
\begin{equation}
    \langle R(t)\rangle \; \stackrel{t\to 0}{=} \; \frac{v_\infty\Gamma_K}{2}t^2,
\end{equation}
and late-time behavior,
\begin{equation}
    \langle R(t)\rangle \; \stackrel{t\to \infty}{=} \; v_\infty\left(t-\frac{1}{\Gamma_K}\right),
\end{equation}
depend on the expansion speed $v_\infty$ in addition to the nucleation rate $\Gamma_K$. In principle, fitting both the early- and late-time predictions for $\langle R\rangle$ could extract the two unknown parameters $v_\infty$ and $\Gamma_K$. However, this is complicated by the modifications to the late-time behavior due to the finite size of the trap [Fig.~\ref{fig:7}(b)] and the noise sensitivity of the initial profile. Measuring only the late-time scaling, as done in \cite{zenesini2024false}, captures the expansion speed in our model, but not the nucleation rate.}

\edit{
Instead, we propose the relaxation of the field in the central region (\mbox{$x=0$}) as an alternative observable that is directly related to the nucleation rate. Indeed, from Eq.~\eqref{eq:1_11_theory} we have
\begin{equation}
    \langle \theta\rangle_{x=0} = \theta\ms e^{-\Gamma_K t} + \theta_{\rm s} \left(1-e^{-\Gamma_K t}\right).
\end{equation}
The value of the metastable and stable phases may be extracted from the initial and final states of an experiment and compared to the theoretical predictions. Because nucleation has been experimentally observed to occur first near \mbox{$x=0$}, the local relaxation in this region isolates the nucleation rate from the subsequent domain expansion.}

\edit{Finally, we note that the fraction of realizations with nucleated domains in experiments \cite{zenesini2024false} is not consistent with the functional form predicted by Eq.~\eqref{eq:nucleatedfraction}. This discrepancy could arise from difficulties in identifying critical droplets with low contrast between interior and exterior [Fig.~\ref{fig:1}(c2)], or from dynamical effects neglected by our overdamped description, such as the inertia of the fields. In our model the linear increase of $R$ is a consequence of the overdamped limit. In the underdamped limit, the constant force \mbox{$\sim\delta V$} acting on the domain walls would yield a parabolic trajectory. Thus, the asymptotic behavior of $R(t)$ may be used to determine whether experiments are in the overdamped or underdamped regime.
}

\section{Conclusions and outlook}
\label{sec:conclusions}

In this paper, we have calculated the nucleation rate of a 1D ferromagnetic superfluid, including fluctuation contributions that set its magnitude. 
Computing these fluctuation corrections for a ferromagnetic superfluid is a priori not an obvious task, even in one dimension, as a straightforward application of an instanton description yields divergent results (cf.~App.~\ref{app:fdkt}).
Our results elucidate the exact parametric dependence of different contributions to the nucleation rate, which is important for a precise comparison between experimental results and theoretical predictions. 

We dedicated particular attention to the small-tilt regime, where we obtained closed-form solutions for the critical-droplet configuration and the fluctuation determinant, and confirmed them against numerically exact results. This regime is ideal for a comparison between theory and experiments, as the condition \mbox{$\beta\Delta E\gg 1$} for the validity of the saddle-point approximation is fulfilled. Moreover, in this regime critical droplets are strongly polarized [Fig.~\ref{fig:1}(e2)], which allows for a clear contrast in experimental images of the density imbalance. Conversely, critical droplets in the small-barrier regime are small and shallow (i.e., the polarization never deviates much from the metastable state, and in particular does not approach the stable-phase polarization) [Fig.~\ref{fig:1}(c2)], which likely makes them hard to distinguish from the metastable background in experimental images. In the small-barrier limit, the field-theoretical description of nucleation in terms of instantons and their fluctuations breaks down, and instead the system undergoes spinodal decomposition.

We have discussed the identification of the nucleation rate from the average domain growth profile, pointing out the existence of two distinct time scales: the formation rate of critical droplets and their rate of expansion. To analyze their different signatures, we extended our results for the small-tilt regime into a model for the collective dynamics of domain walls. This provides a basis to simplify the full field-theoretical nucleation problem to a Kramers escape problem for the size of the nucleating domain. Using this effective theory, we model the average field profile over many realizations of a nucleation experiment, both in a homogeneous background and a trapping potential. Our results indicate that the nucleation rate may be extracted directly from the average population imbalance at the trap center. This metric avoids the timescale mixing and trap-induced inhomogeneities that obscure the effect of the nucleation rate in other observables, such as the total nucleated fluid fraction. We suggest a model for the domain size that includes both the nucleation rate and the \edit{expansion rate} as independent parameters. Because the nucleation rate is controlled by the energy barrier while the \edit{expansion rate} is determined by the stable-state energy, independently tuning these features of the potential should enable a direct confirmation of our predictions.

\edit{Our work is based on the assumptions of a thermal description of nucleation. To estimate the impact of quantum fluctuations on the thermal field theory, we examine the coherent-state action for a Rabi-coupled two-component BEC in the rotating-wave approximation. Quantum corrections are largest in the small-tilt regime, where neglecting fluctuations in fields other than the relative density and requiring that the Matsubara modes give a negligible contribution to the metastable partition function, yields (reinstating units for clarity):
\begin{equation}
    T_{\rm quantum} \lesssim \frac{\hbar|\kappa|n}{2 \pi k_B} \sqrt{1 - \left( \frac{\Omega_R}{|\kappa|n} \right)^2}.
\end{equation}
The parameters from \cite{zenesini2024false} result in \mbox{$\hbar|\kappa|n/(2\pi k_B) \approx 10\,$}nK at the trap center, whereas~\cite{cominotti2025observation} reports temperatures on the order of $\mu$K; this justifies the use of classical thermal nucleation theory. It would be interesting to extend our analysis to include the relative density and phase fields to determine the conditions under which quantum fluctuations would be observable.}

Several open questions remain. We have taken the mean-field energy~\eqref{eq:energy} as the starting point of our analysis, but a formulation including quantum or higher-order effects may be better suited for modeling nucleation experiments. An improvement in this direction is to include the inhomogeneous density profile of the gas directly in the energy functional. This could clarify why nucleation starts at the trap center, a result observed both in experiments and simulations \cite{billam19, brown2025mitigating}. Our methods for the evaluation of the nucleation rate extend directly to these cases, as they apply to 1D energy functionals with general potentials and field-dependent stiffness terms. With the inclusion of a renormalization prescription, they also generalize to higher dimensions \cite{garcia2024fluctuation}. Another direction of improvement concerns the ansatz used in the small-tilt regime. Here, higher-order corrections to the critical-droplet profile could follow from a double expansion in the shape of the domain wall and droplet size, providing more precise results while keeping analytically tractable expressions. Finally, it would be interesting to derive the diffusion and dissipation coefficients used in our effective theory for droplet nucleation and growth in Sec.~\ref{sec:Kramers} from microscopic principles. In particular, a noise strength that increases with density would provide a physical mechanism for the preferred nucleation at the center of the trap.

\section*{Acknowledgements}
We thank Jeff Maki for discussions and comments on the manuscript.

\paragraph{Funding information}
This work is supported by Vetenskapsr\aa det (Grant Nos.~2020-04239 and 2024-04485), the Olle Engkvist Foundation (Grant No.~233-0339), the Knut and Alice Wallenberg Foundation (Grant No.~KAW 2024.0129), and Nordita.

\begin{appendix}
\numberwithin{equation}{section}

\section{Fluctuation determinant from direct diagonalization}
\label{app:direct_diagonalization}

In this appendix, we use direct diagonalization to compute the fluctuation determinant~\eqref{eq:D}, which is evaluated using the Gel'fand-Yaglom method in the main text. Results obtained using direct diagonalization and the Gel'fand-Yaglom method agree, which provides a check for our calculations. This appendix also serves to illustrate the structure of the excitation spectrum and to showcase the efficiency of the Gel'fand-Yaglom method.

As discussed in the main text, the functional determinant of a differential operator may be evaluated as a product of eigenvalues, just like its finite-dimensional counterpart. In this form, Eq.~\eqref{eq:D} may be written as the limit \mbox{$D(a,c) = \lim_{\Lambda\to\infty} D_\Lambda(a,c)$}, where
\begin{equation}
    \ln D_\Lambda(a,c) = \ln|\lambda_{-1}| + \int_{0^+}^{\Lambda} d\lambda \; \ln\lambda \, \left[\frac{dN\cb}{d\lambda} - \frac{dN\ms}{d\lambda}\right]. \label{eq:det_op}
\end{equation}
Here, the logarithm turns the product of eigenvalues into an improper integral with a cutoff~$\Lambda$, and $N\cb(\lambda)$, $N\ms(\lambda)$ are the cumulative densities of states of the critical droplet and metastable state, respectively. The lower integration limit is $0^+$ to omit the zero mode, and we separate the single negative mode~$\lambda_{-1}$ in the spectrum of the critical droplet [cf. Eq.~\eqref{eq:D}]. 

To evaluate Eq.~\eqref{eq:det_op}, we determine the eigenvalues of the fluctuation operator~\eqref{eq:K} for the critical droplet and the metastable state with Dirichlet boundary conditions on $[-L/2, L/2]$. The eigenvalues of $\mathcal{K}\cb$ are generally not known analytically; we compute them numerically, using \texttt{ApproxFun.jl} to discretize the differential operator into a matrix eigenproblem. By contrast, the eigenvalues of $\mathcal{K}\ms$ are the same as for a quantum-mechanical particle in a box, \mbox{$\lambda_n = \alpha n^2 + \beta$} with \mbox{$\alpha = \pi^2/L^2$} and \mbox{$\beta=\partial^2 V/\partial \theta^2|_{\theta\ms}$}. The sum of the first $N$ eigenvalues is given by
\begin{equation}
\int_{0}^{\Lambda} d\lambda \ln\lambda \frac{dN\ms}{d\lambda} = \sum_{n=1}^{N} \ln \lambda_n = N\ln \alpha + \ln\frac{\Gamma\left(N+1+i\sqrt{\beta/\alpha}\right)\Gamma\left(N+1-i\sqrt{\beta/\alpha}\right)}{\Gamma\left(1+i\sqrt{\beta/\alpha}\right)\Gamma\left(1-i\sqrt{\beta/\alpha}\right)} ,
\label{eq:fvSum}
\end{equation}
where the cutoff is linked to the number of eigenvalues by \mbox{$N=\sqrt{(\Lambda-\beta)/\alpha}$}.

\begin{figure}[t]
    \centering
    \includegraphics[width=\linewidth]{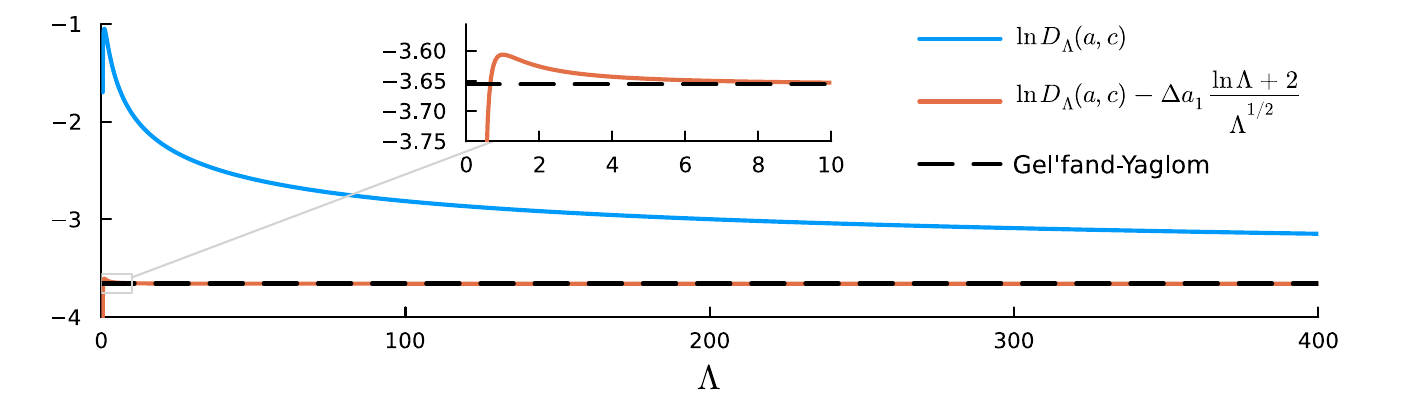}
    \caption{Convergence of the fluctuation determinant~\eqref{eq:det_op} as a function of the cutoff~$\Lambda$ for \mbox{$a=0.2$}, \mbox{$c=0.4$}, and \mbox{$L=60$}. To obtain the blue curve, we sum the logarithm of the eigenvalues of $\mathcal{K}\cb$ up to $\Lambda$, and subtract Eq.~\eqref{eq:fvSum} from the result. \edit{Adding to $\ln D_\Lambda$ the leading $\Lambda$-dependence of the cutoff tail, Eq.~\eqref{eq:logdetHC_cp1}, greatly improves the convergence to the limiting value (orange line).} Both curves coincide in the \mbox{$\Lambda\to\infty$} limit with the Gel'fand-Yaglom result (black dashed line).}
    \label{fig:8}
\end{figure}

The blue line in Fig.~\ref{fig:8} shows \mbox{$\ln D_\Lambda(a=0.2,c=0.4)$} as a function of the cutoff~$\Lambda$, obtained using the numerically computed fluctuation spectrum in combination with the analytical result~\eqref{eq:fvSum}. The result from the Gel'fand-Yaglom method is indicated by a black dashed line. Because Eq.~\eqref{eq:fvSum} is defined for all $N$, we use it to numerically take the limit in Eq.~\eqref{eq:det_op} by evaluating $\ln D_\Lambda$ at the eigenvalues of $\mathcal{K}\cb$. As is apparent from the figure, the convergence to the \mbox{$\Lambda\to\infty$} limit is very slow. Pushing the sum to higher values of $\Lambda$ is computationally expensive (for example, the largest cutoff value \mbox{$\Lambda =400$} in the figure already corresponds to \mbox{$N\approx460$} eigenvalues). \edit{However, we may significantly accelerate the convergence by adding an asymptotic approximation to the difference \mbox{$\ln D -\ln D_\Lambda$} between the limiting value and a finite-$\Lambda$ value (orange line in Fig.~\ref{fig:8}).} The derivation of this limit is discussed in the remainder of this section.

To this end, consider the asymptotic \mbox{$\lambda\to\infty$} cumulative density of states of a general operator \mbox{$-d^2/dx^2 + U(x)$} with Dirichlet boundary conditions, which is given by the Bohr-Sommerfeld quantization rule~\cite{berry1972semiclassical}
\begin{equation}
    N(\lambda) \sim \frac{1}{\pi}\int_{-L/2}^{L/2}dx\,\sqrt{\lambda - U(x)} - \frac{1}{2}. \label{eq:WKB}
\end{equation}
This expression generalizes the quantization condition of a particle in a box, \mbox{$(N + 1/2) \pi = k L$}, by replacing $kL$ with the phase accumulated by the local wave vector \mbox{$k(x) = \sqrt{\lambda-U(x)}$}. We expand the first term in Eq.~\eqref{eq:WKB} in powers of $\lambda$ to obtain
\begin{equation}
    \frac{1}{\pi}\int_{-L/2}^{L/2}dx\,\sqrt{\lambda-U(x)} = a_0\lambda^{1/2} + a_1\lambda^{-1/2} + \mathcal{O}(\lambda^{-3/2}),
\end{equation}
where 
\begin{equation}
    a_0 = \frac{L}{\pi}, \qquad
    a_1 = -\frac{1}{2\pi}\int_{-L/2}^{L/2}dx\,U(x).
\end{equation}
The $a_0$ term is the same for both $N\cb$ and $N\ms$, and gives no contribution to the integral~\eqref{eq:det_op}. \edit{Thus, the leading contribution to the cutoff tail of Eq.~\eqref{eq:det_op} as \mbox{$\Lambda\to\infty$} is given by
\begin{equation}
    \edit{\ln D(a,c) - \ln D_\Lambda(a,c) \sim - \frac{\Delta a_1}{2} \int_\Lambda^\infty d\lambda\,\frac{\ln\lambda}{\lambda^{3/2}} = -\Delta a_1 \frac{\ln\Lambda + 2}{\Lambda^{1/2}},} \label{eq:logdetHC_cp1}
\end{equation}
where $\Delta a_1$ stands for the difference of $a_1$ between $\mathcal{K}\cb$ and $\mathcal{K}\ms$. As we see in Fig.~\ref{fig:8}, adding the contribution of the tail significantly improves the convergence as \mbox{$\Lambda\to\infty$} to the Gel'fand-Yaglom value.} We have systematically checked a \mbox{$20\times 20$} grid of points in our parameter space, and obtained agreement between the Gel'fand-Yaglom and exact diagonalization results with a relative error below $1\%$.

\section{Numerical solution of the Gel'fand-Yaglom equation}
\label{app:numeric_GY}

In this appendix, we describe a method for the precise numerical evaluation of the fluctuation determinant~\eqref{eq:GY_det}, which avoids the \mbox{$\lambda\to0$} limit in favor of a differential equation for the linear-in-$\lambda$ term of $\mathcal{P}\cb(\lambda)$ directly.

Let \mbox{$\phi(x, \lambda) = u\cb(x, \lambda)/u\ms(x, 0)$}, where $u_{\rm sp}(x, \lambda)$ solves Eq.~\eqref{eq:GY_ODE}. We then have \cite{dunne2005beyond}
\begin{equation}
    -\frac{d^2\phi}{dx^2}- 2\sqrt{V''\ms} \coth \left[\sqrt{V''\ms} \left(x+\frac{L}{2}\right)\right]\frac{d\phi}{dx} + \left[V''\cb - V''\ms\right]\phi = \lambda \phi, 
\end{equation}
where we have used the exact solution for $u\ms(x,0)$ and \mbox{$V''_{\rm sp}=\partial_\theta^2V|_{\theta_{\rm sp}}$}. The boundary conditions are given by \mbox{$\phi|_{-L/2}=1$} and \mbox{$d\phi/dx|_{-L/2}=0$}. We then expand \mbox{$\phi = \phi_0 + \lambda \phi_1 + \mathcal{O}(\lambda^2)$} to obtain
\begin{align}
    -\frac{d^2\phi_0}{dx^2}- 2\sqrt{V''\ms} \coth \biggl[\sqrt{V''\ms} \biggl(x+\frac{L}{2}\biggr)\biggr]\frac{d\phi_0}{dx} + \biggl[V''\cb - V''\ms\biggr]\phi_0 &= 0, \\[0.5ex]
    -\frac{d^2\phi_1}{dx^2}- 2\sqrt{V''\ms} \coth \biggl[\sqrt{V''\ms} \biggl(x+\frac{L}{2}\biggr)\biggr]\frac{d\phi_1}{dx} + \biggl[V''\cb - V''\ms\biggr]\phi_1 &= \phi_0 \label{eq:b3}
\end{align}
with boundary conditions
\begin{equation}
     \phi_0\left(-L/2\right)=1, \quad \frac{d\phi_0}{dx}\left(-L/2\right)=0, \quad \phi_1\left(-L/2\right)=0, \quad \frac{d\phi_1}{dx}\left(-L/2\right)=0.
\end{equation}
From Eq.~\eqref{eq:GY_det}, we have \mbox{$D(a,c) = \phi_1(L/2)$}. Compared to Eq.~\eqref{eq:GY_ODE}, which has solutions that diverge exponentially as $x$ increases, the solutions to Eq.~\eqref{eq:b3} remain bounded and quickly converge to the correct value.

\section{Energy functionals with field-dependent stiffness}
\label{app:fdkt}

In this appendix, we compute the fluctuation determinant for energy functionals with a field-dependent stiffness. Such a term is not present in the main text, but it does appear when parameterizing the energy functional~\eqref{eq:energy} in terms of the imbalance \mbox{$Z(x) = (n_{\uparrow} - n_\downarrow)/n$}~\cite{zenesini2024false},
\begin{equation}
    E[Z(x)] = \int_{-\infty}^\infty dx\; \left\{\frac{p[Z]}{2}\left(\frac{dZ}{dx}\right)^2 + V[Z]\right\},
\end{equation}
where
\begin{equation}
    p[Z] = \frac{1}{1-Z^2}, \qquad V[Z] = aZ - \frac{1}{2}Z^2 - c\sqrt{1-Z^2}.
\end{equation}
The imbalance $Z(x)$ is related to our polarization field $\theta(x)$ by the nonlinear transformation \mbox{$Z(x) = \sin[\theta(x)]$}, which maps the derivative term $p[Z](dZ/dx)^2$ to the form $(d\theta/dx)^2$. For a general~$p[Z]$, a simplifying transformation is obtained by solving \mbox{$dZ/d\theta=p^{-1/2}$}. Of course, results for the nucleation rate including fluctuation corrections should not---and, as we show here, do not---depend on the parametrization used. However, this equivalence is not straightforward to see since the fluctuation determinant for a theory with a field-dependent stiffness contains an additional divergent piece. As we show here, this divergent piece is canceled by a corresponding divergence in the Jacobian of the path-integral measure \cite{zinn2021quantum},
\begin{align}
    \mathcal{Z} = \int\mathcal{D}[\theta] e^{-\beta E[\theta]} = \int\mathcal{D}[Z] \left|\det \frac{\delta \theta}{\delta Z}\right| e^{-\beta E} = \int\mathcal{D}[Z]  e^{-\beta E + \frac{1}{2} \text{Tr} \ln p}. \label{eq:SZ}
\end{align}
We now show that the saddle-point evaluation of Eq.~\eqref{eq:SZ} in terms of the imbalance $Z(x)$ is equivalent to the result obtained in the main text.

First, since the Jacobian term is subleading in $\beta$ and can be ignored, the instanton contribution is obtained by taking the $Z$-variation of the exponent in Eq.~\eqref{eq:SZ}, which gives
\begin{equation}
    \frac{\delta E}{\delta Z} = -\left(\sqrt{p[Z]}\frac{d}{dx}\right)^2Z + \frac{\partial V}{\partial Z} = 0. \label{eq:Zeom}
\end{equation}
This is the same instanton equation as obtained when changing variables in Eq.~\eqref{eq:instanton_equation} from $\theta(x)$ to $Z(x)$. Second, expanding \mbox{$E[Z_{\rm sp}(x) + \delta Z(x)]$} in $\delta Z$, we obtain the following second-order contribution:
\begin{equation}
    E[Z_{\rm sp} + \delta Z] = E_{\rm sp} + \frac{1}{2} \int_{-L/2}^{L/2} dx \; \delta Z(x) \, \tilde{\mathcal{K}}_{\rm sp} \, \delta Z(x) + \mathcal{O}(\delta Z^3), 
\end{equation}
where, after some algebra, the fluctuation operator in Sturm-Liouville form reads
\begin{equation}
    \tilde{\mathcal{K}}_{\rm sp} = -\frac{d}{dx}\left(p[Z_{\rm sp}]\frac{d}{dx}\right) -\frac{d^2 p[Z_{\rm sp}]}{dx^2} + \frac{1}{2} \left(\frac{d Z_{\rm sp}}{dx}\right)^2 \frac{\partial^2 p}{\partial Z^2}\Biggr\rvert_{Z_{\rm sp}} + \frac{\partial^2 V}{\partial Z^2}\Biggr\rvert_{Z_{\rm sp}}. \label{eq:KZ}
\end{equation}
Following the same steps as in the main text, the contribution of fluctuations around a critical point $Z_{\rm sp}$ of the energy reads
\begin{equation}
    \mathcal{Z} = e^{-\beta E[Z_{\rm sp}]} \times \exp\left(\frac{1}{2}\text{Tr}\ln p[Z_{\rm sp}]\right) \times \sqrt{\frac{\beta E[Z_{\rm sp}]}{2\pi}} \times |\det{'}\tilde{\mathcal{K}}_{\rm sp}|^{-1/2}.
    \label{eq:sstone}
\end{equation}
Here, the second term results from the Jacobian factor evaluated at the critical point. The next term is the contribution of the zero mode, whose normalization differs from that of $\theta(x)$ because of the nonlinear field redefinition [cf.~Eq.~\eqref{eq:zero_mode}]. The last term is the functional determinant characterizing finite fluctuations of $Z_{\rm sp}$. The Jacobian and determinant factors in Eq.~\eqref{eq:sstone} combine to give
\begin{equation}
    e^{-\text{Tr} \ln p[Z_{\rm sp}]} \det \tilde{\mathcal{K}}_{\rm sp} = \det p^{-1} \det \tilde{\mathcal{K}}_{\rm sp} = \det\left(p^{-1/2} \tilde{\mathcal{K}}_{\rm sp} p^{-1/2}\right) = \det\left(\mathcal{K}_{\rm sp}\right),
\end{equation}
where $\mathcal{K}_{\rm sp}$ is the main-text fluctuation operator~\eqref{eq:K}, and we assume we may factor functional determinants. This completes the equivalence between the one-loop partition function~\eqref{eq:Zratio} using $\theta(x)$ and $Z(x)$.

\end{appendix}

\bibliography{bib_nucleation}

\end{document}